\documentclass[12pt,a4paper]{article}
\usepackage{amsmath,amssymb,graphicx,hyperref}
\usepackage{cite}
\numberwithin{equation}{section}
\allowdisplaybreaks[3]

\begin{document}
\begin{titlepage}
		
\renewcommand{\thefootnote}{\fnsymbol{footnote}}
\begin{flushright}
\begin{tabular}{l}
YITP-17-92\\
\end{tabular}
\end{flushright}
		
\vfill
\begin{center}
			
			
\noindent{\large \textbf{Correlators  in higher spin AdS$_3$ holography}
			
\medskip
			
\noindent{\large \textbf{from Wilson lines with loop corrections} }}
			
\vspace{1.5cm}

\noindent{Yasuaki Hikida$^{a}$\footnote{E-mail: yhikida@yukawa.kyoto-u.ac.jp} and Takahiro Uetoko$^b$\footnote{E-mail: rp0019fr@ed.ritsumei.ac.jp}}

\bigskip
			
\vskip .6 truecm
			
\centerline{\it $^a$Center for Gravitational Physics, Yukawa Institute for Theoretical Physics,}
\centerline{\it  Kyoto University, Kyoto 606-8502, Japan}
\medskip 
\centerline{\it $^b$Department of Physical Sciences, College of Science and Engineering,} 
\centerline{\it Ritsumeikan University, Shiga 525-8577, Japan}

\end{center}
		
\vfill
\vskip 0.5 truecm

\begin{abstract}

We study the correlators of the 2d W$_N$ minimal model in the semiclassical regime with large central charge from bulk viewpoint by utilizing open Wilson lines in $\text{sl}(N) $ Chern-Simons gauge theory.
We extend  previous works for the tree level of bulk theory to incorporate loop corrections in this paper. We offer a way to regularize divergences associated with loop diagrams such that three point functions with two scalars and a higher spin current agree with the values fixed by the boundary W$_N$ symmetry.
With the prescription, we reproduce the conformal weight of the operator corresponding to a bulk scalar up to the two loop order for explicit examples with $N=2,3$.

\end{abstract}
\vfill
\vskip 0.5 truecm
		
\setcounter{footnote}{0}
\renewcommand{\thefootnote}{\arabic{footnote}}
\end{titlepage}
	
\newpage
	
\tableofcontents

\section{Introduction}

In \cite{Hikida:2017byl} we computed three point functions with two scalar operators and a higher spin current in the 2d W$_N$ minimal model with $1/N$ corrections. The main aim of this paper is to give a bulk interpretation of the conformal field theory results.%
\footnote{After completing this draft, we become aware of an interesting paper \cite{Giombi:2017hpr} appearing in the arXiv. The paper deals with loop corrections in two point Witten diagrams for higher spin theories on AdS$_d$. Related previous works may be found in \cite{Manvelyan:2008ks,Hikida:2015nfa,Creutzig:2015hta,Hikida:2016wqj,Hikida:2016cla,Aharony:2016dwx,Hikida:2017ecj,Cardona:2017tsw}.}
The $1/N$ corrections  (or $1/c$ corrections with $c$ as the central charge) in the minimal model should be interpreted as loop corrections in the bulk gravity description. However, it is notoriously difficult to deal with divergences associated with gravitational loop diagrams in general. Applying  holography, it is expected that boundary theory can define bulk quantum theory of gravity generically. For our case, the minimal model would determine the way to regularize these gravitational divergences, and we would like to show that this is indeed the case in this paper.

The 2d W$_N$ minimal model has a coset description as 
\begin{align}
	\frac{\text{su}(N)_k \oplus \text{su}(N)_1 }{ \text{su}(N)_{k+1} } 
	\label{coset}
\end{align}
with the central charge
\begin{align}
	c = (N-1) \left( 1 - \frac{N(N+1)}{(k+N) (k+N+1)}\right) \, .
	\label{central}
\end{align}
In \cite{Gaberdiel:2010pz} the 't~Hooft limit with large $N$ but finite $\lambda = N/(k+N)$ of the minimal model is conjectured to be dual to the classical 3d Prokushkin-Vasiliev theory of \cite{Prokushkin:1998bq}. 
Instead of the 't~Hooft limit, we consider the semiclassical regime with large $c$ but finite $N$. The bulk description for the semiclassical regime is supposed to be given by Chern-Simons gauge theory based on $\text{sl}(N) \oplus \text{sl} (N)$ dressed by perturbative matters \cite{Castro:2011iw,Gaberdiel:2012ku,Perlmutter:2012ds}.
The large $c$ regime should be realized with a negative level $k = -1 - N + \mathcal{O}(c^{-1})$, thus the conformal field theory is non-unitary in the regime.%
\footnote{The analysis of this paper will not rely on unitarity, so we can safely work in the non-unitary regime. However, we may have to make use of unitarity for other purposes, and in that case we should come back to the 't~Hooft limit, for instance, by utilizing the analytic continuation discussed in \cite{Gaberdiel:2012ku}. \label{nonunitary}}
In \cite{Hikida:2017byl} we evaluated correlators at the 't~Hooft limit with $1/N$ corrections, but the results can be generalized for the semiclassical limit with $1/c$ corrections.
We try to interpret the $1/c$ corrections in terms of $\text{sl}(N) $ Chern-Simons gauge theory.

The W$_N$ symmetry of the minimal model is generated by higher spin currents $J^{(s)} (z)$ with $s=2,3,\ldots,N$. We examine the following two and three point functions as
\begin{align}
	\langle  \mathcal{O}_{h_+} (z_1) \bar{\mathcal{O}}_{h_+} (z_2) \rangle \, , \quad
	\langle \mathcal{O}_{h_+} (z_1) \bar{\mathcal{O}}_{h_+} (z_2)  J^{(s)} (z_3) \rangle 
	\label{2&3pt}
\end{align}
including $1/c$ corrections. Here  $\mathcal{O}_{h_+} $   is a scalar operator with  conformal weight $h_+ =  (1 - N)/2 + \mathcal{O}(c^{-1})$. The negative value of the conformal weight reflects the non-unitarity of the theory.
At the leading order in $1/c$, it was claimed in \cite{Besken:2016ooo} that correlators or conformal blocks can be computed by the networks of open Wilson lines in $\text{sl}(N) $ Chern-Simons gauge theory.%
\footnote{Previously, Wilson lines in $\text{sl}(N) $ Chern-Simons gauge theory were utilized to compute entanglement entropy in a holographic way \cite{deBoer:2013vca,Ammon:2013hba}. For the case with $N=2$, the proposal reduces to that in \cite{Ryu:2006bv,Ryu:2006ef}.}
For instance, the expectation value of an open Wilson line computes the two point function $\langle  \mathcal{O}_{h_+}  \bar{\mathcal{O}}_{h_+} \rangle$. 
Roughly speaking, the open Wilson line corresponds to a particle running in the bulk, which is dual to the boundary two point function. 
Furthermore, the three point function $\langle \mathcal{O}_{h_+}  \bar{\mathcal{O}}_{h_+}  J^{(s)}  \rangle$  can be evaluated with the extra insertion of the boundary current $J^{(s)}$.
The main aim of this paper is to interpret the $1/c$ corrections of the correlators \eqref{2&3pt} as loop corrections in the bulk computations with open Wilson lines.
For $N=2$, the Chern-Simons theory reduces pure gravity theory as in \cite{Achucarro:1987vz,Witten:1988hc}, and in that case $1/c$ corrections have been examined in Virasoro conformal blocks \cite{Fitzpatrick:2016mtp} and the conformal weight of the scalar operator \cite{Besken:2017fsj}.
The validity of the method with $N=2$ is formally supported by the analysis of conformal Ward identity \cite{Verlinde:1989ua,Fitzpatrick:2016mtp}. See also \cite{Anand:2017dav} for a recent application.

During loop computations with open Wilson lines, we would meet divergences and a main issue in this paper is to propose a prescription to regularize the divergences. There are three main steps in the prescription. Firstly, we have to decide how to introduce a regulator $\epsilon$ to make integrals finite. We adopt a kind of dimensional regularization such that scaling invariance is not broken. Secondly, we have to remove the terms diverging for $\epsilon \to 0$. Here we choose to shift parameters in the open Wilson line since we cannot remove divergences  in the current setup with the shift of parameters in Lagrangian as for usual quantum field theory.
Finally, we have to remove ambiguities arising from $\epsilon$-independent parts in the shift of parameters. We offer a way to fix them so as to be consistent with the W$_N$ symmetry of the minimal model.

It is easy to show that the Wilson line method reproduces the leading order results for correlators in \eqref{2&3pt} with generic $N$. For $1/c$ corrections, we mainly focus on the simplest examples with $N=2$ and $N=3$. 
We find that the three point functions from the Wilson line method  are regularization scheme dependent at the $1/c$ order. Since the three point functions of the minimal model are fixed by the symmetry, we adopt a regularization such that the Wilson line results match the minimal model ones. 
For $N=2$, the authors in \cite{Besken:2017fsj} tried to reproduce the $1/c$ corrections in the conformal weight of the scalar operator from the bulk theory. They succeeded in doing so up to the $1/c$ order since it is regularization independent, but they failed at the $1/c^2$ order due to the regularization issue.
Adopting our prescription for regularization, we succeed in reproducing the $1/c^2$ order corrections of conformal weight both for $N=2$ and $N=3$.

The organization of this paper is as follows;
In the next section, we summarize the results on two and three point functions \eqref{2&3pt} in the 2d W$_N$ minimal model of \eqref{coset} at the semiclassical limit with $1/c$ corrections.
In  section \ref{Preliminaries}, we explain our prescription to compute boundary correlators in terms of open Wilson lines in sl$(N)$ Chern-Simons gauge theory. We reproduce the minimal model results at the leading order  in $1/c$ and describe our prescription to regularize divergences arising from loop diagrams.
In section \ref{Correlators2}, we apply our method to the simplest case with $N=2$. 
In particular, we reproduce the result in \cite{Besken:2017fsj} for the two point function at the $1/c$ order  and improve their argument for the next order in $1/c$ with the help of our analysis for the three point function.
In section \ref{Correlators3}, we proceed to the $N=3$ case and show that our prescription also works for this example.
In section \ref{conclusion}, we conclude this paper and discuss open problems.

\section{W$_N$ minimal model in the semiclassical regime}
\label{CFT3pt}

In this section, we examine the two and three point functions \eqref{2&3pt} of the coset model \eqref{coset} with large $c$ but finite $N$ in $1/c$ expansion. For this purpose we should describe the model in terms of $c,N$ instead of $k,N$ in \eqref{coset}. The parameter $k$ is related to $c,N$ as
\begin{align}
	k = - 1 - N + \frac{N (N^2 -1)}{c} + \frac{N (1-N^2)(1-N^3)}{c^2} + \mathcal{O} (c^{-3}) 
\end{align}
in $1/c$ expansion. Originally $k$ is a positive integer, but here we assume an analytic continuation of $k$ to a real value. See \cite{Gaberdiel:2012ku} for details on the issue.
Using this relation, we can expand physical quantities in $1/c$, and terms at each order depend only on $N$.

The two point function is fixed by the symmetry as
\begin{align}
	\langle \mathcal{O}_{h}  (z) \bar{\mathcal{O}}_{h} (0) \rangle =
	\frac{1}{|z|^{4 h}} \, ,
	\label{2ptcan0}
\end{align}
where $h$ is the conformal weight of the scalar operator $\mathcal{O}_h$.
The overall normalization can be set as $1$ by changing the definition of $\mathcal{O}_{h}$. This implies that the two point function is obtained only from knowledge of the spectrum. 
Throughout the paper, we only focus on the holomorphic sector, thus we may write
\begin{align}
	\langle \mathcal{O}_{h}  (z) \bar{\mathcal{O}}_{h} (0) \rangle =
	\frac{1}{z^{2 h}}
	\label{2ptcan}
\end{align}
instead of \eqref{2ptcan0}.

The spectrum of primary states can be obtained with finite $k,N$ by applying standard methods
like coset construction as in \cite{Bais:1987zk}. The states are labeled as $(\Lambda_+ , \omega; \Lambda_-)$, where $\Lambda_+ , \omega, \Lambda_-$ are the highest weights of $\text{su}(N)_{k}, \text{su}(N)_{1}, \text{su}(N)_{k+1}$, respectively. 
The selection rule determines $\omega$ in terms of $\Lambda_+ , \Lambda_-$, so we may instead use the label $(\Lambda_+ ; \Lambda_-)$. We should take care of the field identification in \cite{Gepner:1989jq} as well. The conformal weight of the state can be obtained by coset construction \cite{Bais:1987zk} or Drinfeld-Sokolov reduction, see, e.g.,  \cite{Bershadsky:1989mf,Bouwknegt:1992wg}. 
For instance, the latter gives the formula
\begin{align}
	h (\Lambda_+ ; \Lambda_-) = \frac{| (k+N+1) (\Lambda_+ + \hat \rho ) - (k + N) (\Lambda_- + \hat \rho)|^2 - \hat \rho^2 }{2 (k + N ) (k+N+1)} \, , 
\end{align}
where $\hat \rho$ is the Weyl vector of $\text{su}(N)$.
According to \cite{Perlmutter:2012ds} (see also \cite{Castro:2011iw} for the original proposal), the state $(0;\Lambda_-)$ corresponds to a conical defect geometry, and the generic state  $(\Lambda_+;\Lambda_-)$ is mapped to the geometry dressed by perturbative matters. In particular, the states $(0;0)$ and $(\text{f};0)$ correspond to the AdS vacuum, and a bulk scalar field on the background.  Here we denote  $\text{f}$ as the fundamental representation.
The conformal weight of the state $(\text{f};0)$ is
\begin{align}
	&h_+ \equiv h(\text{f};0) = \frac{(N-1) (k + 2N + 1)}{2 N (k + N)} \, ,
	\label{hexact}
\end{align}
and we mainly deal with the operator $\mathcal{O}_{h_+} $ corresponding to the state in this paper.

Expanding the conformal weight $h$ in $1/c$ as 
\begin{align}
	h = h_0 + \frac{1}{c} h_1 + \frac{1}{c^2} h_2 + \mathcal{O} (c^{-3}) \, ,
	\label{hexp}
\end{align} 
the two point function becomes
\begin{align}
	\langle \mathcal{O}_{h}  (z) \bar{\mathcal{O}}_{h} (0) \rangle = \frac{1}{z^{2 h_0}} \left[ 1 -   \frac1c 2 h_1 \log (z)  + \frac{1}{c^2} \left( 2  h_1^2 \log ^2 (z) - 2 h_2 \log (z) \right) \right] + \mathcal{O} (c^{-3}) \, . \label{2ptexp}
\end{align}
For the operator $\mathcal{O}_{h_+} $ we have
\begin{align}
	h_0 =  \frac{1-N}{2} \, , \quad
	h_1 = -\frac{\left(N^2-1\right)^2}{2 } \, , \quad
	h_2 = -\frac{(N+1)^2 (2 N (N+1)+1) (N-1)^3}{2} \, , \label{h012}
\end{align}
which is obtained from the expression \eqref{hexact} with finite $k,N$.
The problem will be whether we can reproduce correct the coefficients in front of $\log (z)$ and $\log^2 (z)$ from the bulk viewpoint with open Wilson lines.

We also examine the three point functions in \eqref{2&3pt}. 
In \cite{Hikida:2017byl} we have evaluated the three point functions by decomposing the four point function of $\mathcal{O}_{h_+} $ with Virasoro conformal blocks.
As seen below, we have  effectively decomposed the W$_N$ vacuum block, which is fixed by the W$_N$ symmetry in principle, and this implies that the three point functions can be fixed solely by the symmetry. 
Notice that the three point function with spin two current as
\begin{align}
	\langle \mathcal{O}_{h}  (z_1) \bar{\mathcal{O}}_{h} (z_2)  J^{(2)} (z_3) \rangle
\end{align}
is determined by the conformal Ward identity, and our conclusion may be regarded as a higher spin generalization.

We decompose the following four point function as 
\begin{align}
	G_{++} (z)  &= \langle \mathcal{O}_{h_+} (\infty) \bar{\mathcal{O}}_{h_+} (1) \mathcal{O}_{h_+} (z) \bar{\mathcal{O}}_{h_+} (0) \rangle \, ,
	\label{4pt}
\end{align}
for which the expression with finite $k,N$ is given by \cite{Papadodimas:2011pf}
\begin{align}
	G_{++} (z)  =  |\mathcal{F}_1 (z)|^2 + \mathcal{N}_1 |\mathcal{F}_2 (z)|^2 
	\, .
\end{align}
Here the W$_N$ conformal blocks are
\begin{align}
	\mathcal{F}_1 (z) = z^{- 2 h_+}(1 - z)^{ - 2 h_+ + \frac{k +2N}{k+N}} 
	{}_2 F_1 \left( \frac{k+N+1}{k+N} , - \frac{1}{k+N} ; - \frac{N}{k+N} ; z \right) \, ,\nonumber  \\
	\mathcal{F}_2 (z) = z^{- 2 h_+ + \frac{k +2N}{k+N}}(1 - x)^{ - 2 h_+ } 
	{}_2 F_1 \left( \frac{k+N+1}{k+N} , - \frac{1}{k+N} ;  \frac{2 k + 3N}{k+N} ; z \right) \, ,  
\end{align}
and the relative coefficient is
\begin{align}
	\mathcal{N}_1 = - \frac{\Gamma (\frac{k + 2N -1}{k+N}) \Gamma (\frac{-N}{k+N})^2\Gamma (\frac{2 k + 3N +1}{k+N})}{\Gamma(\frac{- k - 2N - 1}{k+N}) \Gamma (\frac{1-N}{k+N}) \Gamma (\frac{2k+3N}{k+N})^2} \, .
\end{align}
From the leading terms in $z$ expansion, we can read off the conformal weights of the intermediate state.
For $\mathcal{F}_1 (z)$ and $\mathcal{F}_2 (z)$, the intermediate states are found to be the identity and the state $(\text{adj};0)$, respectively. Here adj represents the adjoint representation of sl$(N)$, and the conformal weight of the state is $h (\text{adj};0) = (k +2N)/(k+N)$. This is consistent with the decomposition as $\text{f} \otimes \bar{\text{f}} = 1 \oplus \text{adj}$ with $\bar{\text{f}}$ as the anti-fundamental representation of sl$(N)$.
As discussed in \cite{Hikida:2017byl},  we only need to consider the W$_N$ vacuum block $\mathcal{F}_1 (z) $ in order to obtain the three point functions in \eqref{2&3pt}. Therefore, we  conclude that these three point functions are fixed by W$_N$ symmetry even with finite $k,N$.

We obtain the three point functions with $1/c$ corrections by slightly modifying the analysis in \cite{Hikida:2017byl}.  We decompose the four point function \eqref{4pt} as
\begin{align}
	|z|^{4 h_+} G_{++} (z) = \mathcal{V}_0 (z)+ \sum_{s =3}^\infty (C^{(s)})^2  \mathcal{V}_s (z) 
	+ \cdots \, ,
	\label{cbd}
\end{align}
where $\mathcal{V}_0 (z)$ is the Virasoro vacuum block and $\mathcal{V}_s (z)$ is the Virasoro block of spin $s$ current. The coefficient $C^{(s)}$ is related to the three point function in \eqref{2&3pt} as
\begin{align}
	C^{(s)} =  \frac{\langle \mathcal{O}_{h_+} \bar{\mathcal{O}}_{h_+}  J^{(s)} \rangle }{\langle  J^{(s)}J^{(s)} \rangle^{1/2} } \, .
\end{align}
Since $\mathcal{V}_s (z)$ start to contribute at the order of $1/c$, we expand as
\begin{align}
	C^{(s)} = c^{-1/2} \left[ C^{(s)}_0 + c^{-1} C^{(s)}_1 + \mathcal{O} (c^{-2}) \right] \, .
\end{align}
The relevant part of the four point function \eqref{4pt} can be expanded in $z$ and $1/c$ as 
\begin{align}
	& |z|^{4 h_+} G_{++}(z) \\
	& \quad \sim 1 + \frac{1}{c} \sum_{n=1}^\infty (1 - N^2)
	\left( - \frac{1}{n} + \frac{N \Gamma(N) \Gamma(n)}{\Gamma(N +n)} \right) z^n + \frac{1}{c^2}  \sum_{n = 2}^\infty f_c^{(n)} z^n + \cdots \, , 
	\nonumber
\end{align}
where we have defined
\begin{align}
	& \frac{f_c^{(n)}}{(1 - N^2)^2} = \frac{1}{n} \sum_{l=1}^{n-1} \frac{1}{l} + \frac{\Gamma(n) \Gamma(N) N^2}{\Gamma(N + n)} 
	\left( \sum_{l=0}^{n-1} \frac{N}{N + l} - \frac{1}{n} - 2 - \frac{1}{N} + \frac{1}{1+N} \right) \nonumber \\ & \qquad\qquad\qquad\qquad\qquad - \sum_{l=1}^{n-1} \frac{N\Gamma(N)\Gamma(l)}{(n-l) \Gamma(N+l)} + \left(2N + \frac{1}{1 + N}\right) \frac{1}{n} \, .
\end{align}

Solving the constraint equations from \eqref{cbd}, we find
\begin{align}
	(C^{(s)}_0)^2 = \frac{(1 - N^2) \Gamma(1+N) \Gamma(s-N)}{\Gamma(1 - N) \Gamma(s + N)} \frac{\Gamma(s)^2}{\Gamma(2s-1)}
	\label{3pt0}
\end{align}
for the leading order in $1/c$.
The first few examples are
\begin{align}
	(C^{(2)}_0)^2 = \frac{1}{2} (1-N)^2 \, , \quad 
	(C^{(3)}_0)^2 = \frac{1}{6} \frac{(1-N)^2 (2 - N)}{(2 + N)} \, .
\end{align}
The square of the three point function could be negative for $N \geq 3$, and this is related to the fact that we are working in a non-unitary theory.

Examining the equation \eqref{cbd} at the next order in $1/c$, we can obtain $1/c$ corrections to the three point functions as well. At this order, the constraint equations for $s=3,4,5$ are found to be
\begin{align}
	&f^{(3)}_c = f^{(2)}_c + 2 C_{0}^{(3)}C_{1}^{(3)} \, , \nonumber
	\\
	&f^{(4)}_c = f^{(2)}_c\frac{9}{10} + \frac{(1-N)^2}{8(1+N)^2} + \frac{1-N}{10(1+N)^2} + \frac{1}{50(1+N)^2} + 2 C_{0}^{(3)}C_{1}^{(3)}\frac{3}{2} + 2 C_{0}^{(4)}C_{1}^{(4)}  \, , \nonumber  \\
	&	f^{(5)}_c = f^{(2)}_c\frac{4}{5} + \frac{(1-N)^2}{4(1+N)^2} + \frac{1-N}{5(1+N)^2} + \frac{1}{25(1+N)^2} 	+ 2 C_{0}^{(3)}C_{1}^{(3)}	\frac{12}{7} + 2 C_{0}^{(4)}C_{1}^{(4)}\cdot2 \nonumber \\
	& \qquad  + 2 C_{0}^{(5)}C_{1}^{(5)}  + (C_{0}^{(3)})^2\left[\frac{1}{2}\frac{1-N}{1+N}+\frac{6}{7(1+N)}+\frac{18}{49(1-N^2)}\right]  \, . \end{align}
From these equations, we obtain
\begin{align}
	&\frac{C_{1}^{(3)}}{C_{0}^{(3)}} = N^3+3 N^2-3 N-\frac{6}{N+2}+1 \, , \nonumber \\
	&	\frac{C_{1}^{(4)}}{C_{0}^{(4)}} = N^3+\frac{29 N^2}{4}+\frac{3 N}{2}+\frac{189}{2 (N-3)}-\frac{8}{N-2}+\frac{47}{40 (N-1)}-\frac{3}{10 (N-1)^2} \nonumber \\ & \qquad  -\frac{27}{40 (N+1)}-\frac{3}{10 (N+1)^2}-\frac{6}{N+2}-\frac{36}{N+3}+\frac{161}{4} \, , \\
	&	\frac{C_{1}^{(5)}}{C_{0}^{(5)}} = N^3+\frac{155 N^2}{12}+\frac{29 N}{2}+\frac{800}{N-4}-\frac{180}{N-3}+\frac{25}{7 (N-1)}
	-\frac{25}{7 (N+1)}-\frac{6}{N+2} \nonumber \\ & \qquad -\frac{36}{N+3}-\frac{120}{N+4}+\frac{359}{2} \, . \nonumber 
\end{align}
In particular, $C_{1}^{(3)}/C_{0}^{(3)} = 224/5$ for $N=3$.
It is not difficult to extend the analysis for $C_{1}^{(s)}/C_{0}^{(s)}$ at least up to $s=8$ by directly applying the analysis in \cite{Hikida:2017ecj}.

\section{Preliminaries for bulk computations}
\label{Preliminaries}

In this section, we explain our prescription to compute the two and three point functions \eqref{2&3pt} from bulk theory. In the next subsection, we introduce sl$(N)$ Chern-Simons gauge theory and open Wilson lines. 
In subsection \ref{slNgen} we explain the representation of sl$(N)$ generators in terms of $x$-derivatives.
In subsection \ref{Tree}, we compute the two and three point functions in \eqref{2&3pt} at the leading order in $1/c$. 
In subsection \ref{Reg}, we give a prescription to regularize divergences arising from loop diagrams, and prepare for explicit computations for $N=2,3$ in succeeding sections.

\subsection{Chern-Simons gauge theory and open Wilson lines}

In three dimensions, pure gravity  with a negative cosmological constant can be described by 
$\text{sl}(2) \oplus \text{sl}(2)$ Chern-Simons gauge theory \cite{Achucarro:1987vz,Witten:1988hc}.
As a natural extension, we can construct a higher spin gauge theory using Chern-Simons theory based on a higher rank gauge algebra \cite{Blencowe:1988gj}.
We are interested in $\text{sl}(N) \oplus \text{sl}(N)$ Chern-Simons theory, whose action  
is given by
\begin{align}
	S = S_\text{CS} [A] - S_\text{CS} [\tilde A] \, , \quad
	S_\text{CS} [A] = \frac{\hat k}{4 \pi} \int \text{tr} \left( A \wedge d A + \frac{2}{3} A \wedge A \wedge A \right) \, .
\end{align}
Here $\hat k$ is the level of Chern-Simons theory and $A, \tilde A$ are one forms taking values in $\text{sl}(N)$. 
The generators of sl$(N)$  can be decomposed in terms of the adjoint action of embedded sl$(2)$ as
\begin{align}
	\text{sl} (N) = \text{sl} (2) \oplus  \left( \bigoplus_{s=3}^{N} g^{(s)} \right) \, .
\end{align}
Here $g^{(s)}$ denotes the spin $(s-1)$ representation of sl$(2)$, and we have adopted the principal embedding of sl$(2)$. The generators in sl$(2)$ (adjoint representation) and $g^{(s)}$  are denoted as  $V^2_n$ $(n=-1,0,1)$ and $V^s_n$ $(n= - s+1,-s+2,\ldots , s-1)$, respectively.

For the application to higher spin AdS$_3$ gravity, we need to assign an asymptotic AdS condition to the gauge fields. We use the metric of Euclidean AdS$_3$ as $ds^2 = d \rho^2 + e^{2\rho} dz d \bar z$, where the boundary is at $\rho \to \infty$. In a gauge choice, we can set
\begin{align}
	A = e^{- \rho V_0^{2}} a (z) e^{\rho V_0^{2}} dz + V_0^2 d \rho \, .
\end{align}
We have a similar expression for $\tilde A$ but suppress it here and in the following.
The configuration corresponding to AdS$_3$ background is given by $a(z) = V_{1}^2$.
The asymptotic AdS condition restricts the form of $a(z)$ as \cite{Henneaux:2010xg,Campoleoni:2010zq,Gaberdiel:2011wb,Campoleoni:2011hg} 
\begin{align}
	a (z) = V_1^{2} - \frac{1}{\hat k} \sum_{s \geq 2} ^N \frac{1}{N_s} J^{(s)} (z) V^s_{-s + 1} \, , \quad N_s = \text{tr} (V_{-s +1}^s V_{s-1}^s) \, .
	\label{dsgauge}
\end{align}
There are residual gauge symmetries preserving the condition \eqref{dsgauge}, and a part of them generates W$_N$ symmetry near the AdS boundary.
We can define classical Poisson brackets  for the reduced phase space. Moreover, we can see that $J^{(s)}(z)$ in \eqref{dsgauge} generate the W$_N$ symmetry in terms of the Poisson brackets.
At the classical level, the relation between the Chern-Simons level $\hat k$ and the central charge $c$ of the dual conformal field theory is given by the Brown-Henneaux one as \cite{Brown:1986nw}
\begin{align}
	c = 6 \hat k \, .
	\label{bhclassical}
\end{align}
See \cite{Henneaux:2010xg,Campoleoni:2010zq,Gaberdiel:2011wb,Campoleoni:2011hg} for more details.

At the leading order in $1/c$, the rules for computing conformal blocks from the Chern-Simons theory with open Wilson lines were given in \cite{Besken:2016ooo}, see also \cite{Bhatta:2016hpz} for $N=2$. For the two and three point functions in \eqref{2&3pt}, we use
\begin{align}
	\langle \text{lw} | W (z_2 ; z_1) | \text{hw} \rangle \, , \quad
	W(z_2 ; z_1) = P \exp \left(\int _{z_1}^{z_2} dz a(z) \right) \, .
	\label{classicalwilson}
\end{align}
Here hw and lw denote the highest and lowest weight states in finite dimensional representations of sl$(N)$, respectively, and $P$ represents the path ordering.
Moreover, we remove the $\rho$-dependence in the gauge field as $A(z) = a(z)$ using  a  gauge transformation. 
We include $1/c$ corrections by extending the analysis in \cite{Fitzpatrick:2016mtp,Besken:2017fsj} for $N=2$. At the leading order in $1/c$, we treat the coefficient $J^{(s)}(z)$ in \eqref{dsgauge} as a function of $z$. At higher orders in $1/c$, we regard $J^{(s)}(z)$ as an operator, and the expectation values of open Wilson lines are evaluated by using the correlators of $J^{(s)}(z)$, which are uniquely fixed by the W$_N$ symmetry.

\subsection{Generators of $\text{sl}(N)$ algebra}
\label{slNgen}

In this subsection we explain our prescription to compute the matrix elements of sl$(N)$ algebra for evaluating the expectation values of open Wilson lines as in \eqref{classicalwilson}.
We start with the simplest case with $N=2$ and then extend the argument for generic $N$.
For $N=2$, there are several previous works in \cite{Verlinde:1989ua,Fitzpatrick:2016mtp,Besken:2017fsj}, and we start  by clarifying the representation with $x$-derivatives in \cite{Fitzpatrick:2016mtp}.

For two point functions we evaluate 
\begin{align}
	\langle j , - j | W_{-j} (z_2 ; z_1) | j , j \rangle \, ,
	\label{Gjz1z2}
\end{align} 
where $| j , m \rangle$ belongs to the spin $j$ representation of sl(2) with $m= -j , - j+1 , \ldots ,j$.
We set the norm of these states as
\begin{align}
	\langle j , m|  j , m' \rangle = \delta_{m,m'} \, .
\end{align}
With these states, the sl(2) generators in the Wilson line are described by $(2j+1) \times (2j+1)$ matrices.

As in \cite{Verlinde:1989ua,Fitzpatrick:2016mtp}, it would be convenient to map the expression as  
\begin{align}
	\langle j , - j | W_{-j} (z_2 ; z_1) | j , j \rangle  = 
	\int dx
	\langle j , - j |x \rangle W_{-j} (z_2 ; z_1) \langle x | j , j \rangle \, ,
\end{align} 
then the sl$(2)$ generators can be written as
\begin{align}
	J_+ (= V_{-1}^2)= x^2 \partial_x - 2 j x \, , \quad 
	J_3 (= - V_{0}^2)= - x \partial_x + j \, , \quad 
	J_-  (= V_{+1}^2)=  \partial_x \, .
	\label{xsl2}
\end{align}
In \cite{Fitzpatrick:2016mtp}, they proposed that the wave functions are given by
\begin{align}
	\langle x | j , j \rangle  = x^{2j} \, , \quad \langle j , - j |x  \rangle = \delta (x) \, .
	\label{basisx}
\end{align}
We would like to give a derivation such that it can be extended for generic $N$.
It is easy to obtain $\langle x | j , j \rangle  = x^{2j}$ as a solution to the equation
$J_+ | j , j \rangle = 0$. The others follow as
\begin{align}
	\langle x |j , m \rangle \propto (J_-)^{j-m}  \langle x | j , j \rangle = \frac{\Gamma(2j+1)}{\Gamma(j+m+1)} x^{j + m} \, .
\end{align}
The dual states $\langle j , m' |x  \rangle $ should satisfy
\begin{align}
	\int dx
	\langle j , m' |x \rangle \langle x | j , m \rangle  = \delta_{m,m'} \, ,
\end{align}
which leads to
\begin{align}
	\langle j , m' |x  \rangle \propto \partial_x^{j + m'} \delta (x) \, .
\end{align}
In particular, we have $\langle j , - j |x  \rangle = \delta (x) $ as in \eqref{basisx}.
The normalization is set to be a convenient value.

We then apply the analysis to the case with generic $N$.
A way to represent the generators of $\text{sl}(N)$ is using $N \times N$ matrices,
and sl(2) generators  $V^{2}_{n}$ $(n=-1,0,1)$ can be embedded as described, e.g., in appendix A of \cite{Castro:2011iw}.
Then the other generators may be obtained as
\begin{align}
	V^{s}_n = (-1)^{s-1-n} \frac{(n+s-1)!}{(2s - 2)!} [V_{-1}^2 [V_{-1}^2 , ... , [ V_{-1}^2 , (V_{1}^2)^{s-1} ]]] \, ,
	\label{slNgenerators}
\end{align}
where $(s - n - 1)$ of  $V_{-1}^2$ are inserted. The fundamental representation of sl$(N)$ can be described by an $N$ dimensional vector, which behaves as a spin $(N-1)/2$ representation under the action of the embedded sl$(2)$. Therefore, the description with $N \times N$ matrices can be given by \eqref{Gjz1z2} with $j= (N-1)/2$ and open Wilson lines based on sl$(N)$ algebra.
In this specific case, we can map the matrix representation to the one with $x$-derivatives using \eqref{xsl2} and \eqref{slNgenerators}. In the representation with $x$-derivatives, the generators of sl$(N)$  should be given by \cite{Bergshoeff:1991dz}
\begin{align}
	V^s_n = \sum_{i=0}^{s-1} (n - s + 1)_{s - 1 - i} a^{i} (s , h_0) x^{- n+i} \partial_x^{i} \, ,
	\label{generators0}
\end{align}
where
\begin{align}
	\quad	a^i (s , h_0) = \binom{s-1}{i} \frac{(- 2 h_0 - s + 2)_{s - 1 - i}}{(s + i)_{s - 1 -i}}  
	\label{generators}
\end{align} 
with $h_0 = - j = (1 - N)/2$. The wave functions are precisely those in \eqref{basisx}.
The generators \eqref{generators0} with \eqref{generators} are those of higher spin algebra hs$[\lambda]$ for $h_0 = (1 + \lambda)/2 $, and 
sl$(N)$ can be realized by hs$[-N]/\chi_N$ with $\chi_N$ as an ideal, which removes generators with $s > N$.

With the realization of generators, $N_s$ in \eqref{dsgauge} are computed as
\begin{align}
	N_s 
	= \frac{3 \sqrt{\pi }  \Gamma (s) (1-N)_{s-1} (N+1)_{s-1}}{ 2^{2s-2} \left(N^2-1\right) \Gamma \left(s+\frac{1}{2}\right)} \, ,
\end{align}
where the first few expressions are 
\begin{align}
	N_2 = -1 \, , \quad N_3 = \frac{1}{5} (N^2 - 4) \, ,\quad N_4 = - \frac{3}{70} (N^2 - 4) (N^2 - 9) \, . 
\end{align}
In particular, we have $N_3 = 1$ for $N=3$.

\subsection{Correlators at the leading order in $1/c$}
\label{Tree}

In order to compute the correlators in \eqref{2&3pt},
we need to consider the expectation values of open Wilson lines with $| \text{hw} \rangle$
corresponding to the highest weight in the fundamental representation of sl$(N)$. 
As explained above, they can be expressed for $(z_1,z_2) = (0,z)$ as
\begin{align}
	W_{h_0} (z)  &= \int dx \delta (x)   P \exp \left[ \int^{z}_{0} d z ' \left( V_1^{2} - \frac{1}{\hat k} \sum_{s = 2}^N \frac{1}{N_s} J^{(s)} (z ') V^s_{-s + 1} \right) \right]  \frac{1}{x^{2h_0}} \nonumber \\
	&= \left.  P \exp \left[ \int^{z}_{0} d z ' \left( V_1^{2} - \frac{1}{\hat k} \sum_{s = 2}^N \frac{1}{N_s} J^{(s)} (z ') V^s_{-s + 1} \right) \right]  \frac{1}{x^{2h_0}} \right |_{x=0}
	\label{Wilson}
\end{align}
with $h_0= (1-N)/2$.
Here the $\text{sl}(N)$ generators are written in terms of $x$-derivatives as in \eqref{generators}.
We would like to treat them perturbatively in $1/\hat k$ (or $1/c$).
Following the analysis in \cite{Besken:2017fsj}, we compute
\begin{align}
	\frac{d}{dz} \left[e^{-z \partial_x } W_{h_0} (z) \right]
	= \left( - \frac{1}{\hat k} \sum_{s = 2}^N \frac{1}{N_s} J^{(s)} (z) e^{- z\partial_x }  V^s_{-s + 1} e^{ z \partial_x } \right) \left[e^{- z\partial_x } W_{h_0} (z) \right] \, .
\end{align}
Integrating over $z$, we find
\begin{align}
	\label{Wilson1/c}
	W_{h_0} (z) &  = \sum_{n=0}^\infty \left( - \frac{1}{\hat k} \right )^n 
	\int_0^z dz_n \cdots \int_0^{z_2} dz_1 \sum_{s_j = 2}^N
	\left[ \prod_{j=1}^n \frac{1 }{N_{s_j}} J^{(s_j)} (z_j)  \right]
	f_n^{(s_n,\ldots , s_1)} (z_n ,\ldots , z_1) \, , 
\end{align}
where
\begin{align}
	\label{fnss}
	&f_n^{(s_n,\ldots , s_1)} (z_n ,\ldots , z_1) \\
	& \qquad = \left.
	\prod_{j=1}^n \left[ 
	\sum_{i=0}^{s_j-1} ( - 2 s_j + 2)_{s_j - 1 - i} a^{i} (s_j , h_0) (x + z - z_j)^{s_j - 1+i} \partial_x^{i}  \right]  \frac{1}{(x + z)^{2h_0}} \right |_{x=0} \, , \nonumber 
\end{align}
see (3.3) of \cite{Fitzpatrick:2016mtp} for $N=2$.

According to the current prescription, the two point function of $\mathcal{O}_{h_+}$ in \eqref{2&3pt} should be computed as
\begin{align}
	\langle \mathcal{O}_{h_+} (z) \bar{\mathcal{O}}_{h_+} (0) \rangle =  \langle W_{h_0} (z) \rangle  \, ,
	\label{2ptWilson}
\end{align}
where  $\langle W_{h_0} (z) \rangle$ is evaluated by the correlators of $J^{(s)}$ in  the  W$_N$ theory. 
The leading order expansion in $1/\hat k$ leads to
\begin{align}
	\left.	\langle \mathcal{O}_{h_+} (z) \bar{\mathcal{O}}_{h_+} (0) \rangle \right|_{\mathcal{O} (c^0) }= \left.	\langle W_{h_0} (z) \rangle \right|_{\mathcal{O} (c^0) }= 
	\frac{1}{z^{2h_0}}
\end{align}
as expected.

We are also interested in the three point functions in \eqref{2&3pt}, which should be obtained as
\begin{align}
	\langle \mathcal{O}_{h_+} (z) \bar{\mathcal{O}}_{h_+} (0) J^{(s)} (y) \rangle =
	\langle  W_{h_0} (z) J^{(s)} (y) \rangle \, .
	\label{3ptWilson}
\end{align}
The first non-trivial contributions come from the terms of order $1/\hat k$ . 
At this order, we need to compute
\begin{align}
	\left.	\langle  W_{h_0} (z) J^{(s)} (y) \rangle  \right|_{\mathcal{O} (c^{0}) }  &=  - \frac{1}{\hat k N_{s}}  
	\int_0^z  dz_1 f^{(s)}_1 (z_1) \langle J^{(s)} (z_1) J^{(s)} (y) \rangle \\
	&    =  - \frac{1}{\hat k N_{s}}  
	\int_0^z  dz_1  \frac{\Gamma (2h _0+ s - 1)}{\Gamma (2 h_0)}\frac{ (z - z_1)^{s-1} z_1^{s-1}}{ z^{ s -1 + 2h_0}}  \langle J^{(s)} (z_1) J^{(s)} (y) \rangle \, . \nonumber
\end{align}
The normalization of higher spin currents in \eqref{dsgauge} corresponds to (see, e.g., \cite{Ammon:2011ua})
\begin{align}
	\left.	\langle J^{(s)} (z_1) J^{(s)} (z_2) \rangle \right|_{\mathcal{O} (c) } = - (2 s -1) \hat k N_s \frac{1}{z_{12}^{2s}} \, .
	\label{2ptnorm}
\end{align}
Using
\begin{align}
	\int_0^z dz_1 \frac{(z - z_1)^{s-1} z_1^{s-1}}{(z_1 - y)^{2s} } 
	= \frac{z^{2s -1}}{( y- z)^s y^s} \frac{(\Gamma(s))^2}{\Gamma (2s)} \, ,
\end{align}
we find 
\begin{align}
	\left.	\langle  W_{h_0} (z) J^{(s)} (y) \rangle   \right|_{\mathcal{O} (c^{0}) }   =  \frac{\Gamma (2h_0 + s - 1)}{\Gamma (2 h_0)}
	\frac{(\Gamma(s))^2}{\Gamma (2s - 1)} 
	\left( \frac{z}{(y-z) y} \right)^s \left.	\langle W_{h_0} (z) \rangle \right|_{\mathcal{O} (c^0) } \, . \label{3pttree}
\end{align}
The result is consistent with \eqref{3pt0} in the convention of \eqref{2ptnorm}.
In fact, it is the same as eq.~(1.3) of \cite{Ammon:2011ua} up to a factor if we set $h_0 = (1 + \lambda)/2$ (or $N = - \lambda$), and this is related to the triality relation discussed in \cite{Gaberdiel:2012ku}.

\subsection{Prescription for regularization}
\label{Reg}

The $1/c$ corrections of the  two and three point functions in \eqref{2&3pt} can be evaluated from higher order contributions in \eqref{Wilson1/c} using the Wilson line method. However, integrals over $z_j$ diverge when two (or more) currents $J(z_i)$ collide. Therefore, we need to decide how to deal with these divergences, and we explain our prescription in this subsection.

Let us start with the correlators of higher spin currents, which are uniquely fixed by the W$_N$ symmetry in terms of central charge $c$. In particular, we use the two point functions 
\begin{align}
	\langle J^{(s)} (z_2) J^{(s)} (z_1) \rangle  = - \frac{ (2 s -1) c N_s}{6} \frac{1}{z_{21}^{2s}} \, ,
	\label{2ptnormc}
\end{align}
which reduce to \eqref{2ptnorm} if we use the relation $c=6 \hat k$ in \eqref{bhclassical}.
At finite $\hat k$, the relation of \eqref{bhclassical} should be modified, and corrections to higher spin propagators are automatically included by expanding in $1/c$ instead of $1/\hat k$, see \cite{Besken:2017fsj} for some arguments. 
Divergence would arise at the coincident point $z_2 = z_1$, and we need to decide how to regularize it.
We introduce a regulator as
\begin{align}
	\langle J^{(s)} (z_2) J^{(s)} (z_1) \rangle  = - \frac{ (2 s -1) c N_s}{6} \frac{1}{z_{21}^{2s - 2 \epsilon}} 
	\label{2ptnormreg}
\end{align}
by shifting the conformal weight of the higher spin current as $s \to s - \epsilon$. 
This choice is reasonable since it does not break the scaling symmetry.
Analogously, we introduce the regulator $\epsilon$ to other correlators of higher spin currents $J^{(s)}$
by shifting the conformal wights of the current.

Introducing the regulator $\epsilon$, integrals over $z_j$ become finite but have terms diverging at $\epsilon \to 0$. In the usual quantum field theory with a renormalizable Lagrangian, we can remove divergences by renormalizing the overall normalization of quantum fields and the parameters of interactions.
In the current case, we offer to remove  divergences in a similar manner.
We first use the fact that the normalization of a two point function can be chosen arbitrarily by the redefinition of the operator. We remove a kind of divergence by changing the overall factor of the open Wilson line such that the corresponding two point function becomes the normalized one as in \eqref{2ptcan}. 
We then notice that the three point interactions between two scalars and a higher spin field are
governed by the coefficients in front of $J^{(s)} (z)$ in \eqref{Wilson}. 
We introduce parameters $c_s$ such that \eqref{Wilson} becomes
\begin{align}
	W_{h_0} (z)  = \left.  P \exp \left[ \int^{z}_{0} d z ' ( V_1^{2} - \frac{6}{c} \sum_{s =2}^N \frac{c_s}{N_s} J^{(s)} (z ') V^s_{-s + 1} ) \right]  \frac{1}{x^{2h_0}} \right |_{x=0} \, .
	\label{Wilsonreg}
\end{align}
In terms of $1/c$ expansion, \eqref{Wilson1/c} is changed as 
\begin{align}
	\label{Wilson1/creg}
	W_{h_0} (z) &  = \sum_{n=0}^\infty \left( - \frac{6}{c} \right )^n 
	\int_0^z dz_n \cdots \int_0^{z_2} dz_1 \sum_{s_j = 2}^N
	\left[ \prod_{j=1}^n \frac{c_{s_j}}{N_{s_j}} J^{(s_j)} (z_j)  \right]
	f_n^{(s_n,\ldots , s_1)} (z_n ,\ldots , z_1) \, , 
\end{align}
where $f_n^{(s_n,\ldots , s_1)} (z_n ,\ldots , z_1) $ are given by \eqref{fnss}.
At the leading order in $1/c$, $c = 6 \hat k$ as in \eqref{bhclassical} and $c_s = 1$.
From the next order in $1/c$, we shift the values of $c_s$ to remove divergences.
Namely, we expand $c_s$ in $1/c$ as
\begin{align}
	c_s = 1 + \frac{1}{c} c_s^{(1)} + \frac{1}{c^2} c_s^{(2)} + \mathcal{O} (c^{-2}) \, ,
	\label{csexp}
\end{align}
and absorb divergences in $c_s^{(i)}$ $(i=1,2,\ldots)$ order by order.
We conjecture that all divergences can be removed by these two ways of renormalization.

As explained above, we determine to remove divergences by properly choosing the ``bare'' values of parameters $c_s$. However, we have still freedom to choose the terms independent of $\epsilon$.
Here we fix them such that the three point functions  $\langle \mathcal{O}_{h_+}  \bar{\mathcal{O}}_{h_+}  J^{(s)}  \rangle$ in \eqref{2&3pt} are reproduced from the Wilson line method as in \eqref{3ptWilson}. Since the three point functions can be fixed by the W$_N$ symmetry as shown in the previous section, we would say that the regularization scheme is determined by making use of the boundary symmetry. This is expected to fix all the ambiguities left, and other physical quantities should be predictable.
In the following two sections, we examine concrete examples with $N=2,3$ and show that the $1/c$ corrections in the conformal dimensions of scalar operators can be reproduced from the bulk viewpoint up to the two loop level applying the prescription described above.

\section{Correlators for $N=2$}
\label{Correlators2}

In this and the next section, we explicitly evaluate the loop corrections of the correlators in terms of open Wilson lines.
We start with the simpler case with $N=2$ and then move to a more involved one with $N=3$.
For $N=2$, we can work with generic $h_0 = -j$, because the sl(2) generators in terms of $x$-derivatives as in \eqref{xsl2} are available for the generic case as argued in subsection \ref{slNgen}.

Two and three point functions with generic $h_0 $ are obtained from analysis of conformal field theory as follows. For $h_0 = - j$, the $1/c$ correction of conformal weight is given as \eqref{hexp} with
\begin{align}
	h_1 = - 6 h_0 (h_0-1) \, , \quad h_2 = - 78 h_0 (h_0-1) \, ,
	\label{dimcorr}
\end{align}
see, e.g., \cite{Besken:2017fsj}.
The $1/c$ expansion of the two point function is then \eqref{2ptexp}.
In the next subsection, we examine the two point function at the next leading order in $1/c$.
We reproduce the order $1/c$ result as $h_1$ in \eqref{dimcorr}, and remove a divergence by renormalizing the overall factor of the open Wilson line. 
The three point function is fixed by the conformal Ward identity as
\begin{align}
	\langle \mathcal{O}_h (z) \bar{\mathcal{O}}_h (0) J^{(2)} (y) \rangle =\left[ h_0 + \frac1c h_1 \right] \left( \frac{z}{(y - z) y} \right)^2  \langle \mathcal{O}_{h} (z) \bar{\mathcal{O}}_{h} (0)  \rangle + \mathcal{O}(c^{-2})
	\label{Ward}
\end{align}
in the  current convention of $J^{(2)}$ given by \eqref{2ptnormc}.
The $c^0$ order term follows from \eqref{3pttree}.
In subsection \ref{3ptN2}, we fix the parameter $c_2$ introduced in \eqref{Wilsonreg} such that the $1/c$ order term  is reproduced. In particular, this removes another type of divergence.
With the regularization scheme, we reproduce the order $1/c^2$ term as $h_2$ in \eqref{dimcorr} from  two point function at the two loop order in subsection \ref{2ptN22}.

\subsection{Two point function at $1/c$ order}
\label{2ptN21}

\begin{figure}
	\centering
	\includegraphics[keepaspectratio, scale=0.7]
	{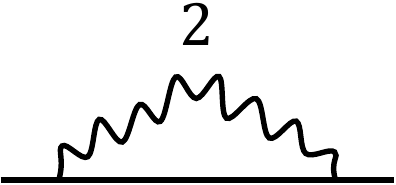}
	\caption{Diagram contributing to the $1/c$ order correction of $\langle \mathcal{O}_h \bar{\mathcal{O}}_h \rangle$ for $N=2$. The straight line and the wavy line represent the open Wilson line and the propagator of spin two current.}
	\label{Wilson1}
\end{figure}

For the two point function of $\mathcal{O}_h$, we need to evaluate the expectation value of the open Wilson line $W_{h_0} (z)$ as in \eqref{2ptWilson}.
With $N=2$, the $1/c$ expansion of the open Wilson line in \eqref{Wilson1/creg} becomes 
\begin{align}
	W_{h_0} (z) =  \frac{1}{z^{2h_0}} + \sum_{n =1} \left( \frac{6 c_2}{c} \right)^n W^{(n)}_{h_0} (z) 
	\label{Wexp}
\end{align}
with
\begin{align}
	&W^{(1)}_{h_0} (z) =  \int_0^z d z_1  f_1^{(2)} ( z_1) J^{(2)}  (z_1) \, , \nonumber \\
	&W^{(2)}_{h_0} (z) = \int_0^z d z_2 \int_0^{z_2} dz_1  f_2^{(2,2)} (z_2,z_1)
	J^{(2)}  (z_2) J^{(2)}  (z_1) \, , \label{Wexpe} \\
	&W^{(3)}_{h_0} (z) = \int_0^z d z_3 \int_0^{z_3} d z_2 \int_0^{z_2} dz_1   f_3^{(2,2,2)} ( z_3,z_2 ,z_1) J^{(2)}  (z_3)J^{(2)}  (z_2)J^{(2)}  (z_1) \, , \nonumber  \\
	&W^{(4)}_{h_0} (z) = \int_0^z d z_4  \cdots \int_0^{z_2} dz_1   f_4^{(2,2,2,2)} ( z_4,z_3 ,z_2,z_1) J^{(2)}  (z_4)J^{(2)}  (z_3) J^{(2)}  (z_2) J^{(2)}  (z_1) \, , \nonumber 
\end{align}
and so on. Here $f_n^{(2,\ldots,2)} (z_n , \ldots , z_1)$ are defined in \eqref{fnss}.
Since the one point function vanishes as $\langle J^{(2)} (z) \rangle = 0$, the non-trivial contribution starts from $\langle W^{(2)}_{h_0} (z) \rangle$. The contribution corresponds to the one loop correction in the two point function of $\mathcal{O}_h$ as in figure~\ref{Wilson1}.

The integrals in $\langle W^{(2)}_{h_0} (z) \rangle$ over $z_1,z_2$ diverge, 
and we introduce a regulator $\epsilon$ as in \eqref{2ptnormreg}, i.e., 
\begin{align}
	\langle J^{(2)} (z_2) J^{(2)} (z_1) \rangle = \frac{c/2}{z_{21}^{4 - 2 \epsilon}} 
\end{align}
for spin two current. 
With the regulator, we obtain a finite result after the integration over $z_1,z_2$ as 
\begin{align}
	&  \langle W^{(2)}_{h_0} (z)  \rangle =  \int_0^z d z_2 \int_0^{z_2} dz_1  f_2 ^{(2,2)} (z_2,z_1)
	\langle J^{(2)}  (z_2) J^{(2)}  (z_1) \rangle \nonumber \\
	& \qquad =  \frac{c}{2 z^{2h_0}} \left[ \frac{(h_0-1) h_0 }{3 \epsilon}+\frac{1}{9} h_0  \left(6 (h_0-1) \log \left(z\right)+5 h_0-2\right) \right] + \mathcal{O} (\epsilon) \, .
	\label{W2}
\end{align}
Using \eqref{Wexp} and $c_2 = 1 + \mathcal{O}(c^{-1})$, the above expression leads to
\begin{align}
	\langle W_{h_0} (z) \rangle = \frac{1}{z^{2h_0}} \left[ 1 + \frac{1}{c}\left(\frac{6(h_0-1) h_0 }{ \epsilon} +  \left(12 h_0 (h_0 -1) \log \left(z\right)+2 h_0 (5 h_0 -2 ) \right) \right)\right] 
	\label{2pt1}
\end{align}
up to the terms of order $\epsilon^0$ and $1/c$.

We compare the above expression in \eqref{2pt1} with the $1/c$ expansion of two point function in \eqref{2ptexp}. 
We can see that the $\log (z)$ term correctly explains $h_1 = - 6 h_0 (h_0-1)$ in \eqref{dimcorr} as shown in \cite{Besken:2017fsj}.
The expression in \eqref{2pt1} has a term proportional to $1/\epsilon$, which diverges for $\epsilon \to 0$.  We can remove the divergence by changing the overall factor of the open Wilson line as
\begin{align}
	\tilde W_{h_0} (z) = \left[ 1 - \frac{1}{c}\left(\frac{6(h_0-1) h_0 }{ \epsilon} +  2 h_0 (5 h_0 -2 )  \right)\right] W_{h_0} (z) \, .
	\label{overall}
\end{align}
With the normalization, we have
\begin{align}
	\langle \tilde W_{h_0} (z) \rangle = \frac{1}{z^{2 (h_0 + h_1/c)}} + \mathcal{O} (c^{-2})  
\end{align}
for $\epsilon \to 0$. In other words, we choose the $\epsilon$-independent part such that the corresponding two point function has unit normalization as in \eqref{2ptcan}.

\subsection{Three point function}
\label{3ptN2}

\begin{figure}
	\centering
	\includegraphics[keepaspectratio, scale=0.7]
	{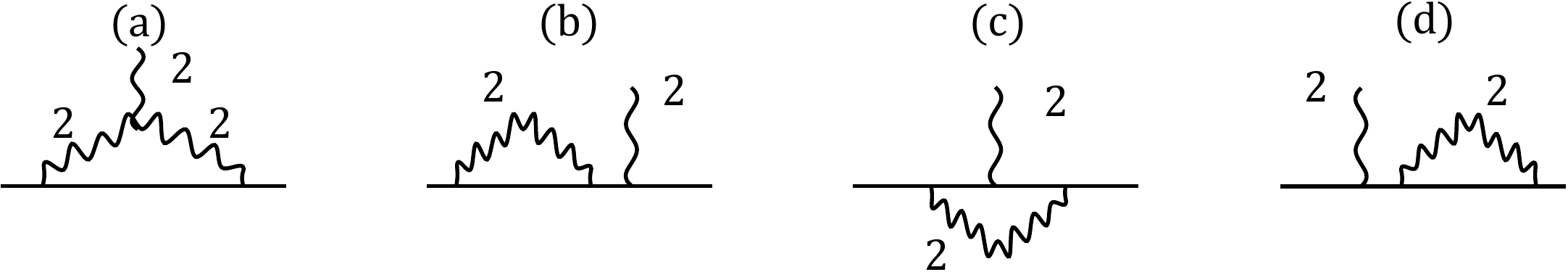}
	\caption{Diagrams contributing to the $1/c$ order correction of $\langle \mathcal{O}_h \bar{\mathcal{O}}_h J^{(2)} \rangle$ for $N=2$.}
	\label{Wilson2}
\end{figure}

We have proposed that three point functions can be computed with open Wilson lines as in \eqref{3ptWilson} and  reproduced the tree level results as in \eqref{3pttree}.
In this subsection, we examine the next leading order in $1/c$.
There are two types of contribution at the order as in figure~\ref{Wilson2}
and we would like to examine them in turn.

The first one is from 
\begin{align}
	\langle W_{h_0}^{(2)} (z) J^{(2)} (y) \rangle  = \int_0^z d z_2 \int_0^{z_2} dz_1  f_2^{(2,2)} (z_2,z_1)
	\langle  J^{(2)}  (z_2) J^{(2)}  (z_1)  J^{(2)} (y)  \rangle \, ,
\end{align}
which is represented as diagram (a) in figure~\ref{Wilson2}.
Here we need to introduce the regulator $\epsilon$ to the three point function of spin two current.
Our prescription is to shift the conformal weight from $2$ to $2 - \epsilon$, so we use
\begin{align}
	\langle  J^{(2)}  (z_2) J^{(2)}  (z_1)  J^{(2)} (y)  \rangle = 
	\frac{c}{z_{21}^{2- \epsilon}(z_2  - y  )^{2- \epsilon}(z_1 - y )^{2- \epsilon} } \, .
\end{align}
The integral becomes simpler by taking $y \to - \infty$ as 
\begin{align}
	\lim_{y \to - \infty} |y|^{4 -2 \epsilon } \langle W_{h_0}^{(2)} (z) J^{(2)} (y) \rangle
	= - \frac{c h_0}{z^{2h_0 - 2}} \left[  \frac{1 }{3 \epsilon} + \frac{1}{36} (18 h_0 +12 \log (z)-13) \right] 
	\label{spin2first}
\end{align}
up to the term of order $\epsilon^0$.

The second one is from
\begin{align}
	&\langle W^{(3)}_{h_0} (z) J^{(2)} (y) \rangle  \nonumber
	\\ & = \int_0^z d z_3 \int_0^{z_3} d z_2 \int_0^{z_2} dz_1   f_3 ^{(2,2,2)} ( z_3,z_2 ,z_1)  \langle J^{(2)}  (z_3)J^{(2)}  (z_2)J^{(2)}  (z_1) J^{(2)} (y) \rangle \, .
\end{align}
At the leading order in $1/c$, the four point function is given by a sum over the products of the two point function as
\begin{align}
	& 	\langle  J^{(2)}  (z_3)J^{(2)}  (z_2)J^{(2)}  (z_1) J^{(2)} (y)  \rangle \nonumber  \\ &=	\frac{c^2/4}{z_{32}^{4 -2 \epsilon}(z_1 - y  )^{4- 2 \epsilon }} +\frac{c^2/4}{z_{31} ^{4 - 2 \epsilon} (z_2 - y )^{4- 2 \epsilon }  }  + 	\frac{c^2/4}{z_{21}^{4- 2\epsilon} (z_3 - y  )^{4- 2 \epsilon}  } 
	+ \mathcal{O}(c)\, . 
\end{align}
Denoting
\begin{align}
	H^{(3)}_{ij} (z) =z^{2h_0 - 2}
	\int_0^z d z_3 \int_0^{z_3} d z_2 \int_0^{z_2} dz_1   f_3 ^{(2,2,2)} ( z_3,z_2 ,z_1)  
	\frac{1}{z_{ji} ^{4 - 2 \epsilon}} \, ,
\end{align}
we find
\begin{align}
	H^{(3)}_{12} (z)& =H^{(3)}_{23} (z)= \frac{(h_0-1) h_0^2 }{9 \epsilon}+\frac{2}{45} h_0  (5 (h_0-1) h_0 \log (z)+(h_0-2) (h_0+1)) \, , \\
	H^{(3)}_{13} (z)& = -\frac{h_0 ((h_0-1) h_0-1) }{9 \epsilon}-\frac{1}{135} h_0  (30 ((h_0-1) h_0-1) \log (z)-h_0 (13 h_0+32)+1)  \nonumber 
\end{align}
up to the terms of  $\mathcal{O} (\epsilon^0)$. The integrals $H^{(3)}_{12}(z)$, $H^{(3)}_{13}(z)$, $H^{(3)}_{23}(z)$ correspond to the diagrams (b), (c), (d) in figure~\ref{Wilson2}, respectively.

Combining the results so far, we find
\begin{align}
	&\lim_{y \to - \infty} |y|^{4 - 2 \epsilon } \langle W_{h_0} (z) J^{(2)} (y)\rangle 
	\nonumber  \\
	&= \frac{1}{z^{2h_0-2}} \left[
	h_0 + \frac{h_0}{c} \left(\frac{6 (h_0(h_0-1)-1) }{\epsilon} +  10 h_0 (h_0-1) + 3 + 12 (h_0-1) h_0 \log ( z) \right) \right] + \cdots \nonumber  \\
	&=  z^2  \left[ h_0 - \frac{6 h_0}{c} \left( \frac{1}{\epsilon} + h_0 - \frac12 \right) \right] \langle W_{h_0} (z) \rangle + \cdots \, .
\end{align}
The expression diverges for $\epsilon \to 0$, and we remove the divergence by properly choosing $c_2^{(1)}$ in \eqref{csexp} as 
\begin{align}
	c_2 =  1 +  \frac{6}{c} \left(  \frac{1}{\epsilon} + a  \right) + \mathcal{O} (c^{-2}) \, .
\end{align}
Here $a$ is an arbitrary constant, which shall be fixed shortly.
With this choice of the parameter $c_2$, there arises a contribution of order $1/c$ from the following term as
\begin{align}
	\left.	\frac{6 c_2}{c} \lim_{y \to - \infty} |y|^{4 - 2 \epsilon } \langle W_{h_0}^{(1)} (z) J^{(2)} (y)\rangle 
	\right|_{\mathcal{O}(c^{-1})}	= \frac{ h_0 }{z^{2h_0 -2}}\frac{6}{c} \left(  \frac{1}{\epsilon} + a \right) 
\end{align}
up to the terms of order $\epsilon^0$. Here $W_{h_0}^{(1)} (z)$ is given in \eqref{Wexpe}.
With this prescription, we have
\begin{align}
	&\lim_{y \to - \infty} |y|^{4 - 2 \epsilon} \langle W_{h_0} (z) J^{(2)} (y)\rangle  \nonumber \\
	& \qquad =  z^2 \left[ h_0 - \frac{6 h_0}{c} \left( h_0 - \frac12 - a \right) \right] \langle W_{h_0} (z) \rangle
	+ \mathcal{O}(c^{-2})
\end{align}
for $\epsilon \to 0$.
Therefore, setting $a = 1/2$, we reproduce the expected result as  \eqref{Ward} with $h_1 = - 6 h_0 (h_0 -1)$ in \eqref{dimcorr}.
In summary, we choose the parameter $c_2$ in \eqref{Wilsonreg} as 
\begin{align}
	c_2 =  1 +  \frac{1}{c} \left(  \frac{6}{\epsilon} + 3  \right) + \mathcal{O} (c^{-2})
	\label{level2c}
\end{align}
in order to absorb a divergence from the one loop diagram and also reproduce the result from the conformal Ward identity.

\subsection{Two point function at $1/c^2$ order}
\label{2ptN22}

In the previous subsections we have regularized divergences arising up to the one loop order.
Our claim is that other quantities are predictable after the renormalization.
Here we would like to examine the two point function at the two loop order.
Generically two loop diagrams have one loop sub-diagrams, and there would appear non-local divergences from the sub-diagrams. After all one loop divergences are removed by  renormalization procedure, we should have no non-local divergences at the two loop order.
There would be local divergences remaining, which can be renormalized as for the one loop computations.
As discussed in \cite{Besken:2017fsj},  two point function without proper renormalization does not reproduce the correct dependence on $\log (z)$ and $\log^2 (z)$ at the two loop order because of non-local divergences as $1/\epsilon \log (z)$.
Since now it is not expected to have such divergences after the renormalization, it should be possible to reproduce the correct shift of conformal weight even at the $1/c^2$ order.
We shall show that this is indeed the case in this subsection.

\begin{figure}
	\centering
	\includegraphics[keepaspectratio, scale=0.7]
	{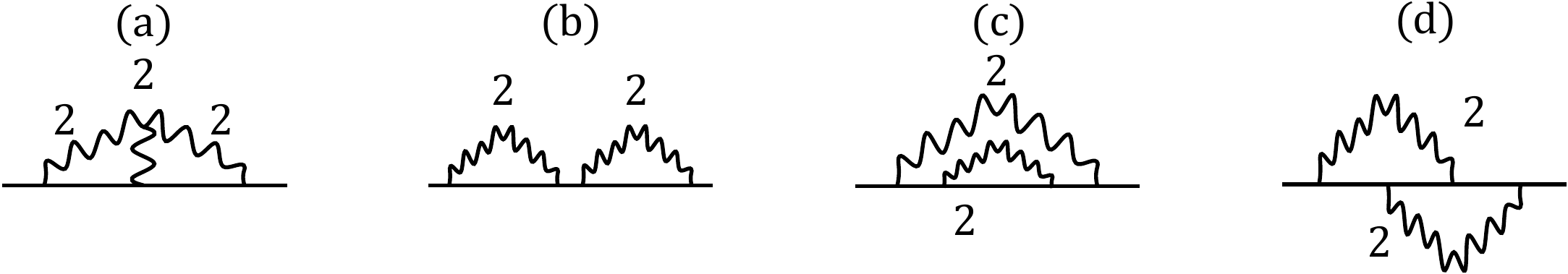}
	\caption{Diagrams contributing to the $1/c^2$ order correction of $\langle \mathcal{O}_h \bar{\mathcal{O}}_h \rangle$ for $N=2$.}
	\label{Wilson3}
\end{figure}

We first evaluate the expectation value of  the open Wilson line at the $1/c^2$ order without renormalization,
then we consider its effects.
A contribution comes from $ \langle W_{h_0}^{(3)} (z) \rangle$ in \eqref{Wexpe} as 
\begin{align}
	G^{(2)}_{123} (z)  = & \left( \frac{6}{c}\right)^3 \int_0^z d z_3 \int_0^{z_3} d z_2 \int_0^{z_2} dz_1   f_3^{(2,2,2)} ( z_3,z_2 ,z_1) 	\frac{c}{z_{32}^{2- \epsilon}z_{31}^{2-\epsilon}  z_{21}^{2- \epsilon}  } \, ,
\end{align}
which is expressed as diagram (a) in figure~\ref{Wilson3}.
The integral is computed as
\begin{align}
	c^2 z^{2h_0}G^{(2)}_{123} (z) 	 = & -\frac{ 288 h_0  (h_0-1)    \log (z)}{\epsilon}  \nonumber \\ &  +2 h_0  \left(36 \log (z) \left(-6 (h_0-1) \log (z)-2 h_0^2-9 h_0+5\right)\right) \, .
\end{align}
Here we neglect the terms of $\mathcal{O}(\epsilon)$ and  write down only the terms depending on $\log (z)$ or $\log ^2 (z)$. In the rest of this subsection, we include only such terms.
Another type of contribution arises from  $ \langle W_{h_0}^{(4)} (z) \rangle$ in \eqref{Wexpe}.
Defining
\begin{align}
	&G^{(2)}_{ij;kl} (z) =  \left( \frac{6}{c}\right)^4 \int_0^z d z_4 \int_0^{z_4} d z_3 \int_0^{z_3} d z_2 \int_0^{z_2} dz_1   f_4^{(2,2,2,2)} ( z_4,z_3,z_2,z_1) 	\frac{c^2/4}{z_{lk} ^{4- 2 \epsilon} z_{ji}^{4- 2 \epsilon}} \, ,
\end{align}
we find
\begin{align}	
	&c^2 z^{ 2 h_0} G^{(2)}_{12;34} (z)  =\frac{144  h_0^2  (h_0-1)^2  \log (z)}{\epsilon} \nonumber \\ & \quad  - 96 (h_0-1) h_0 ^2 \log (z) (-3 (h_0-1) \log (z)-5 h_0+2) \, ,   \nonumber \\
	&c^2 z^{2 h_0} G^{(2)}_{14;23} (z) =  \frac{ 360 (h_0-1)^2 h_0^2  \log (z)}{5 \epsilon} \nonumber \\ & \quad +\frac{72}{5} h_0  (h_0-1) \log (z) (10 (h_0-1) h_0 \log (z)+h_0 (23 h_0-43)-16)  \, ,  \\
	&c^2 z^{2 h_0} G^{(2)}_{13;24} (z) =   - \frac{ 720 h_0 \left((h_0-2) h_0^2+1\right)  \log (z)}{5 \epsilon} \nonumber \\ & \quad +\frac{12}{5} h_0  \log (z) \left(-120 \left((h_0-2) h^2_0 +1\right) \log (z)+h_0 \left(-238 h_0^2+596 h_0+3\right)-241\right)  \, .  \nonumber 
\end{align}
These integrals correspond to diagrams (b), (c), (d) in figure~\ref{Wilson3}, respectively.
Summing over all contributions we find
\begin{align}
	\label{G2z}
	&c^2 z^{2h_0} G^{(2)}_{h_0} (z) =  \frac{ 72 h_0 (h_0-2) (h_0-1) (h_0+1) \log (z) }{\epsilon} \nonumber \\  & \quad +12 h_0  \log (z) \left(12 \left((h_0-2) h_0^2+1\right) \log (z) +h_0 (4 h_0 (5 h_0-7)-5)+1\right) \, . 
\end{align}
Therefore, a non-locally divergent term as $1/\epsilon \log (z)$ remains, and the expression cannot be compared with \eqref{2ptexp}.

Now we include the effects of renormalization, namely, the change of overall factor as in \eqref{overall} and the shift of parameter $c_2$ as in \eqref{level2c}. These effects lead to an extra contribution as
\begin{align}
	\langle \tilde W_{h_0} (z) \rangle 
	&= \left[ 1 - \frac{1}{c}\left(\frac{6(h_0-1) h_0 }{ \epsilon} +  2 h_0 (5 h_0 -2 )  \right) \right] \nonumber \\
	&\times
	\left[ \frac{1}{z^{2h_0}} + \left(\frac{6 }{c} \right)^2  \left( 1 + \frac{1}{c} \left( \frac{6}{\epsilon} + 3 \right) \right)^2  
	\langle W^{(2)}_{h_0} (z) \rangle +  G^{(2)}_{h_0} (z)  \right] + \cdots \nonumber \\
	&= \frac{1}{z^{2h_0}} + \left(\frac{6 }{c} \right)^2\langle W^{(2)}_{h_0} (z) \rangle +  G^{(2)}_{h_0} (z) + \tilde G^{(2)}_{h_0} (z) + \cdots \, , 
\end{align}
where
\begin{align}
	\tilde G^{(2)}_{h_0} (z) = \frac{1}{c}\left[ 2 \left( \frac{6}{\epsilon} + 3 \right) - \frac{6 h_0 (h_0 -1)}{\epsilon} - 2 h_0 (5h_0 -2) \right] \left( \frac{6}{c} \right)^2 \langle W^{(2)}_{h_0} (z) \rangle \, .
	\label{extra}
\end{align}
The extra contribution can be evaluated as 
\begin{align}
	& c^2 z^{2h_0} \tilde G^{(2)}_{h_0} (z) =
	\frac{- 72 h_0  (h_0-2) (h_0-1) (h_0+1) \log \left(z\right)}{\epsilon}   \\
	&  \quad - 72 h_0 (h_0-2) (h_0-1) (h_0+1) \log ^2\left(z\right)  
	+24 h_0 (h_0 (2 h_0 (7-5 h_0)+9)-7) \log \left(z\right)  \nonumber \, .
\end{align}
Thus in total we arrive at
\begin{align}
	c^2 z^{2 h_0} \left. \langle \tilde W_{h_0} (z) \rangle \right|_{\mathcal{O}(c^{0})} = 
	72 h_0^2  (h_0-1)^2  \log ^2 (z)+156 h_0   (h_0-1) \log (z)  \, , 
\end{align}
which does not have any non-local divergence.
Compared with the $1/c$ expansion of two point function in \eqref{2ptexp},
the coefficients in front of $\log (z)$ and $\log^2 (z)$ at the $1/c^2$ order are correctly reproduced with $h_1,h_2$ in \eqref{dimcorr}.

\section{Correlators for $N=3$}
\label{Correlators3}

In the previous section, we have illustrated our prescription by examining a simple example of sl$(N)$ Chern-Simons theory with $N=2$. In this section, we extend the analysis to more involved case with $N=3$.
It is a rather straightforward generalization even though computations become complicated due to the existence of spin three current $J^{(3)}$.
In this paper, we adopt the representation of sl$(N)$ generators with $x$-derivatives as in \eqref{generators}, which is valid for arbitrary representation with $h_0 = - j$ for $N=2$ but only for the fundamental representation with $h_0 = (1 - N)/2$ for $N \geq 3$.%
\footnote{One may find the expression of sl$(3)$ generators for generic representation in terms of three parameters $x_1,x_2,x_3$, e.g., in section 15.7.4 of \cite{cft}.}
With $N=3$,  the $1/c$ expansion of conformal weight is given by \eqref{hexp} with \eqref{h012} as 
\begin{align}
	h_0 = - 1 \, , \quad h_1 = - 32 \, , \quad h_2 = - 1600 \, .
	\label{dimshift3}
\end{align}
In the next subsection, we reproduce the conformal weight at the $1/c$ order as in $h_1$ above from the bulk viewpoint and renormalize open Wilson line. In subsection \ref{3ptN3}, we examine three point functions and fix the two parameters $c_2$ and $c_3$ in \eqref{Wilsonreg} to be consistent with symmetry. In subsection \ref{2ptN32}, we show that our prescription correctly reproduces the conformal weight at the $1/c^2$ order as $h_2$ in \eqref{dimshift3}.

\subsection{Two point function at $1/c$ order}
\label{2ptN31}

\begin{figure}
	\centering
	\includegraphics[keepaspectratio, scale=0.7]
	{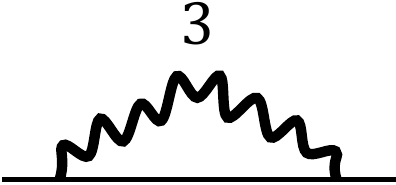}
	\caption{Diagram contributing to the $1/c$ order correction of $\langle \mathcal{O}_{h_+} \bar{\mathcal{O}}_{h_+} \rangle$ for $N=3$ in addition to the one in figure~\ref{Wilson1}. The thick wavy line represents the propagator of spin three current.}
	\label{Wilson4}
\end{figure}

As for $N=2$, we start by examining the two point function at the $1/c$ order.
Since spin three current $J^{(3)}$ is involved along with spin two current $J^{(2)}$,
there are two types of corrections as
\begin{align}
	\left( \frac{6}{c}\right)^2 \left[  \langle W^{(2)}_{h_0} (z) \rangle  + \langle  W^{(2)'}_{h_0 } (z) \rangle \right]  
	\label{W2h}
\end{align}
at this order.	 The two are represents in  figure~\ref{Wilson1} and figure~\ref{Wilson4}, respectively.
Here $W^{(2)}_{h_0} (z) $ is defined in \eqref{Wexpe} and 
\begin{align}
	W^{(2) '}_{h_0} (z) = \int_0^z d z_2 \int_0^{z_2} dz_1 f^{(3,3)}_2 (z_2,z_1) 
	J^{(3)} (z_2) J^{(3)} (z_1)  \, .
	\label{W2p}
\end{align}
Since we have already computed $\langle W^{(2)}_{h_0} (z) \rangle$ as in \eqref{W2}, we just need to evaluate $\langle W^{(2)'}_{h_0} (z) \rangle$.
The prescription in \eqref{2ptnormreg} leads us to adopt
\begin{align}
	\langle J^{(3)} (z_2) J^{(3)} (z_1) \rangle  = - \frac{5c}{6} \frac{1}{z_{21}^{6 -2 \epsilon} }
\end{align}
with the shift of conformal dimension of $J^{(3)}$ from $3$ to $3 - \epsilon$.
Using this expression, we find
\begin{align}
	\langle W^{(2)'}_{h_0} (z) \rangle   = &	- \frac{5 c}{6}\left[
	-\frac{h_0 (h_0 (4 (h_0-2) h_0+1)+3) z^{-2 h_0}}{15 \epsilon} \right. \\ &  \left. 
	-\frac{1}{450} h_0 (2 h_0+1) z^{-2 h_0} (60 (h_0-1) (2 h_0-3) \log (z)+h_0 (94 h_0-115)-9)  \right] \nonumber
\end{align}
up to the term of order $\epsilon^0$. 
Inserting $h_0 = -1$,  we obtain
\begin{align}
	z^{-2}	\langle W_{-1} (z) \rangle= &   1 + \left( \frac{6}{c} \right)^2 \frac{c}{2} \left(\frac{2 }{3 \epsilon} + \frac{1}{9} \left(12 \log \left(z\right) + 7\right) \right)  
	\nonumber \\ &   + \left( \frac{6}{c} \right)^2 \left(- \frac{5c}{6} \right) \left(-\frac{2 }{3 \epsilon}-\frac{1}{450} (600 \log (z)+200) \right) \nonumber \\
	=  &  1 + \frac{1}{c}  \left(\frac{32 }{\epsilon }+ 64 \log (z)+\frac{82}{3} \right)  
\end{align}
up to the terms of orders $\epsilon^0$ and $1/c$.
In particular, the $1/c$ order correction of conformal weight is read off as $h_1 =  - 32$, which is consistent with \eqref{dimshift3}.
In order to remove the divergence at $\epsilon \to 0$ up to the $1/c$ order, we renormalize the Wilson line operator as
\begin{align}
	\tilde W_{-1} (z) = \left[ 1 - \frac{1}{c}  \left(\frac{32}{ \epsilon} +  \frac{82}{3}  \right)  \right]W_{-1} (z) \, , \label{wfren3}
\end{align}
which leads to the corresponding two point function of canonical form as in \eqref{2ptcan}.

\subsection{Three point functions}
\label{3ptN3}

We move to  three point functions with one conserved current. 
For $N=3$, there are two choices of currents, i.e., spin two current $J^{(2)}$ and spin three current $J^{(3)}$.  We start by computing 
$
\langle W_{-1} (z) J^{(2)}(y) \rangle
$
up to the $1/c$ order by following the previous analysis for $N=2$.
With the convention of $J^{(2)}$ in \eqref{2ptnormc}, the corresponding three point function is given by \eqref{Ward} with \eqref{dimshift3}.
At the leading order in $1/c$, we have already computed as in \eqref{3pttree} with $s=2$ and $h_0=-1$.
In the following we shall examine the next non-trivial order in $1/c$.

\begin{figure}
	\centering
	\includegraphics[keepaspectratio, scale=0.7]
	{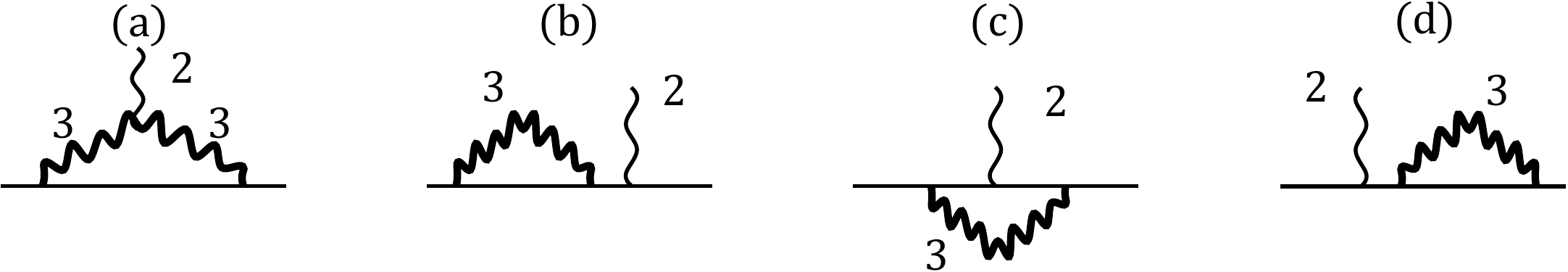}
	\caption{Diagrams contributing to the $1/c$ order correction of $\langle \mathcal{O}_{h_+} \bar{\mathcal{O}}_{h_+} J^{(2)} \rangle $ for $N=3$ in addition to the ones in figure~\ref{Wilson2}.}
	\label{Wilson5}
\end{figure}

At the order in $1/c$, there are several types of contribution as in figure~\ref{Wilson2} and  figure~\ref{Wilson5}.
One  comes from 
\begin{align}
	&\langle W_{-1}^{(2)} (z) J^{(2)}(y) \rangle =  \int_0^z dz_2 \int_0^{z_2} dz_1  f^{(2,2)}_2 (z_2 , z_1)  \langle J^{(2)} (z_2) J^{(2)} (z_1) J^{(2)} (y) \rangle  \, , \nonumber  \\
	&	\langle W_{-1}^{(2)'} (z) J^{(2)}(y) \rangle  =
	\int_0^z dz_2 \int_0^{z_2} dz_1  f^{(3,3)}_2 (z_2 , z_1) \langle J^{(3)} (z_2) J^{(3)} (z_1) J^{(2)} (y) \rangle\, ,  
\end{align}
where $W_{-1}^{(2)} (z)$ and $W_{-1}^{(2)'} (z)$ were introduced in \eqref{Wexpe} and \eqref{W2p}, respectively.
The first contribution corresponds to the diagram (a) in figure~\ref{Wilson2} and it has been computed as in \eqref{spin2first} with $h_0=-1$. 
For the second one corresponding to the diagram (a) in figure~\ref{Wilson5}, we find
\begin{align}
	\lim_{y \to - \infty} |y|^{4 - 2 \epsilon}	\langle W_{-1}^{(2)'} (z) J^{(2)}(y) \rangle  =	- \frac{5  c}{2}  \left[ -\frac{2 z^4}{3 \epsilon} + 
	\frac{1}{18} z^4 (13- 12 \log (z)) \right] + \mathcal{O}(\epsilon) \, ,
\end{align}
where we have used 
\begin{align}
	\langle J^{(3)} (z_2) J^{(3)} (z_1) J^{(2)}(y) \rangle = \frac{- 5  c/2}{z_{21}^{4 - \epsilon} (z_2 - y )^{2 - \epsilon} (z_1 - y )^{2 - \epsilon}   } 
	\label{j3j3j2}
\end{align}
with the shifts of conformal weight both for  $J^{(2)}$ and $J^{(3)}$.

Other types of contribution include four conserved currents.
One of them involves four spin two currents as
\begin{align}
	\int_0^{z} dz_3 \int_0^{z_3} dz_2 \int_0^{z_2} dz_1 
	f^{(2,2,2)}_3 (z_3 , z_2 , z_1 ) \langle J^{(2)} (z_3) J^{(2)} (z_2) J^{(2)} (z_1) J^{(2)} (y) \rangle \, .
\end{align}
They are represented in figure~\ref{Wilson2} and have already been evaluated in subsection \ref{3ptN2}.
Others involve two spin two and two spin three currents, and the correlator of them is factorized at the leading order in $1/c$ as
\begin{align}
	\langle J^{(3)} (z_3) J^{(3)} (z_2)J^{(2)} (z_1)   J^{(2)} (y) \rangle 
	= \frac{- 5c^2/12}{z_{32}^{6 -2 \epsilon}  ( z_1 - y )^{4 - 2 \epsilon}  } + \mathcal{O} (c) \, . \label{j2j3j3j2}
\end{align} 
Therefore, we need to evaluate 
\begin{align}
	&H^{(3,2)}_{1}(z) =   - \frac{5 c^2}{12}  z^{-4} \int_0^{z} dz_3 \int_0^{z_3} dz_2 \int_0^{z_2} dz_1 
	f^{(2,3,3)}_3 (z_3 , z_2 , z_1 ) \frac{1}{z_{21}^{6 - 2\epsilon}  } \, ,\nonumber \\
	&H^{(3,2)}_{2} (z) =   - \frac{5 c^2}{12}  z^{-4}\int_0^{z} dz_3 \int_0^{z_3} dz_2 \int_0^{z_2} dz_1 
	f^{(3,2,3)}_3 (z_3 , z_2 , z_1 ) \frac{1}{z_{31}^{6 - 2\epsilon}  } \, , \\
	&H^{(3,2)}_{3} (z)=  - \frac{5 c^2}{12} z^{-4} \int_0^{z} dz_3 \int_0^{z_3} dz_2 \int_0^{z_2} dz_1 
	f^{(3,3,2)}_3 (z_3 , z_2 , z_1 ) \frac{1}{z_{32}^{6 - 2\epsilon}  } \, , \nonumber 
\end{align}
which correspond to diagrams (b), (c), (d) in figure~\ref{Wilson5}, respectively.
Explicitly performing the integrals, we find
\begin{align}
	& H^{(3,2)}_{1} (z)= H^{(3,2)}_3 (z) =  - \frac{5c^2}{12} \left[\frac{2 }{9 \epsilon}+\frac{1}{9}  (4 \log (z)-1) \right] \, , \nonumber \\
	&H^{(3,2)}_{2} (z)= - \frac{5c^2}{12}  \left[\frac{1}{9 \epsilon}+\frac{1}{54}  (12 \log (z)-13) \right] \, .
\end{align}

Combining  all contributions so far, we have
\begin{align}
	&z^{-4} \lim_{y \to - \infty} |y|^{4 - 2 \epsilon} \langle W_{-1} (z) J^{(2)} (y)\rangle 
	= -1-  \frac{1}{c}    \left[ \frac{6}{\epsilon}+24 \log (z)+23 \right]   \\
	&+ \left(\frac{6}{c} \right)^2 
	\frac{-5 c }{2} \cdot \left[ -\frac{2 }{3 \epsilon} + 
	\frac{1}{18} (13- 12 \log (z)) \right]  +   \left(\frac{6}{c}\right)^3 [2 H^{(3,2)}_1 (z)+ H^{(3,2)}_2 (z)] + \cdots \, . \nonumber
\end{align}
The above expression reduces to
\begin{align}
	- 1 + \frac{1}{c} \left( \frac{4}{\epsilon} - 64 \log (z) - \frac{139}{3} \right)  
	=z^{-2} \left[ -1 +  \frac{1}{c} \left( \frac{36}{\epsilon} -  19  \right)  \right] \langle W_{-1} (z) \rangle   \, .
	\label{wj2}
\end{align}
As before, we choose the parameter $c_2$  in \eqref{Wilsonreg} as
\begin{align}
	c_2 = 1 + \frac{1}{c} \left( \frac{36}{\epsilon} + 13 \right) + \mathcal{O}(c^{-2}) \, .
	\label{level2c32}
\end{align}
This leads to an extra contribution up to the $1/c$ order from
\begin{align}
	\left. 
	z^{-4} \lim_{y \to - \infty} |y|^{4 - 2 \epsilon}  \frac{6c_2}{c} \langle W^{(1)}_{-1} (z) J^{(2)} (y) \rangle 
	\right|_{\mathcal{O}(c^{-1})} = -  \frac{1}{c} \left( \frac{36}{\epsilon} + 13 \right) 
\end{align}
with $W^{(1)}_{h_0}$ in \eqref{Wexpe}.
We can see that this contribution cancels the divergence in \eqref{wj2} as
\begin{align}
	&z^{-4} \lim_{y \to - \infty} |y|^{4 - 2 \epsilon } \langle W_{-1} (z) J^{(2)} (y)\rangle 
	=z^{-2} \left[ -1 -   \frac{32}{c}  \right] \langle W_{-1} (z) \rangle + \mathcal{O}(c^{-2})
\end{align}
for $\epsilon \to 0$.
The constant term in \eqref{level2c32} is chosen in order to reproduce \eqref{Ward} with 
$h_1 = - 32$ as in \eqref{dimshift3}.

We would like to compute another correlator as
$
\langle W_{-1} (z) J^{(3)}(y) \rangle 
$
with spin three current at the $1/c$ order. 
Using the leading order result in \eqref{3pttree} with $h_0 = -1, s=3$,  and the $1/c$ correction as $C_{1}^{(3)}/C_{0}^{(3)} = 224/5$ obtained in section \ref{CFT3pt}, the corresponding three point function is given by
\begin{align}
	\langle \mathcal{O}_{h_+} (z) \bar{\mathcal{O}}_{h_+} (0) J^{(3)} (y) \rangle = \frac{1}{3}\left[ 1 + \frac1c \frac{224}{5} \right] \left( \frac{z}{(y - z) y} \right)^3  \langle \mathcal{O}_{h_+} (z) \bar{\mathcal{O}}_{h_+} (0)  \rangle + \mathcal{O}(c^{-2}) \, .
	\label{3pts3N3}
\end{align}
Following the prescription discussed in subsection \ref{Reg}, we choose the parameter $c_3$ in \eqref{Wilsonreg} such that the Wilson line computation reproduces this expression.

\begin{figure}
	\centering
	\includegraphics[keepaspectratio, scale=0.7]
	{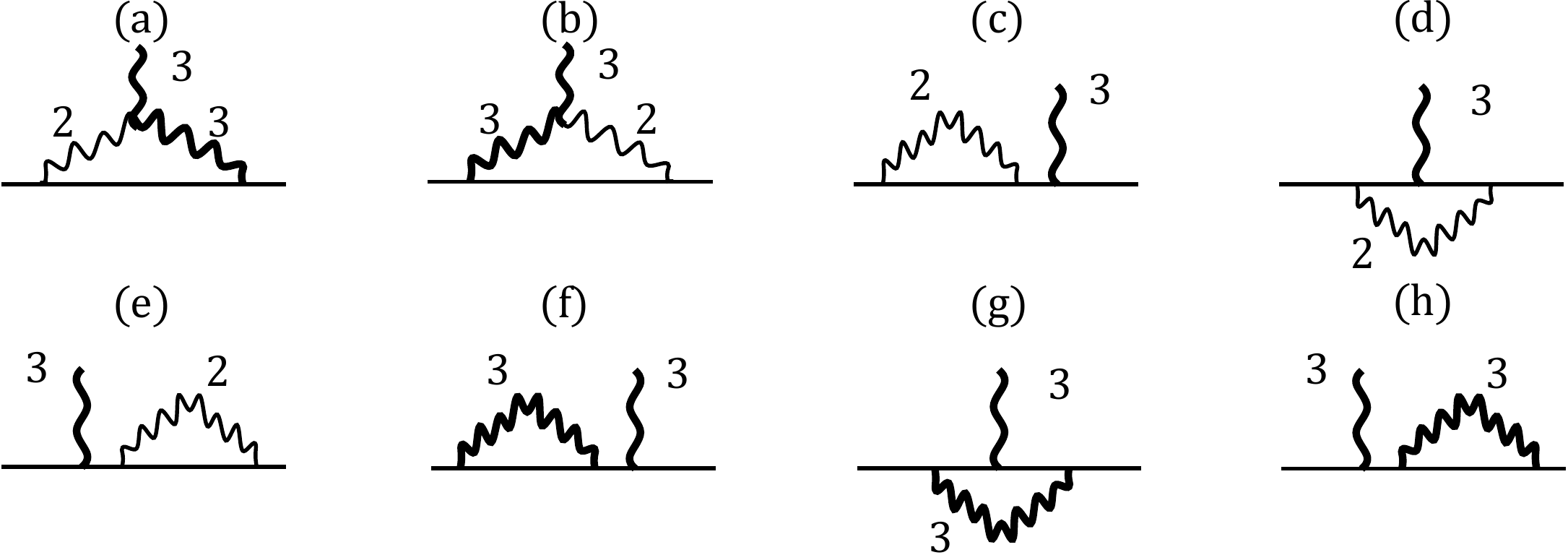}
	\caption{Diagrams contributing to the $1/c$ order correction of $\langle \mathcal{O}_{h_+} \bar{\mathcal{O}}_{h_+} J^{(3)} \rangle $ for $N=3$.}
	\label{Wilson6}
\end{figure}

With two current insertions from an open Wilson line, we have the following type of contribution as
\begin{align}
	H^{(2,3)}	(z)&  =  \lim_{y \to - \infty} |y|^{6  -2 \epsilon} z^{-5} \int_0^z dz_2 \int_0^{z_2} dz_1 \left[  f^{(3,2)}_2 (z_2 , z_1) \right . \langle J^{(3)} (z_2)  J^{(2)} (z_1) J^{(3)} (y) \rangle  \nonumber  \\ & \qquad \qquad \qquad \qquad  + \left.  f^{(2,3)}_2 (z_2 , z_1) \langle  J^{(2)} (z_2)  J^{(3)} (z_1)J^{(3)} (y) \rangle \right] \nonumber \\
	& =- \frac{5 c}{2}  \left[ -\frac{4 }{15 \epsilon} + \frac{1 }{225} (107-60 \log (z)) \right] \, ,
\end{align}
which come from diagrams (a), (b) in figure~\ref{Wilson6}.
For the correlator of three currents, we have used \eqref{j3j3j2}.
There are contributions with two spin two and two spin three currents.  With the correlator in \eqref{j2j3j3j2}, they are given by
\begin{align}
	H^{(3,3)}_{1} (z)=  - \frac{5 c^2 }{12}  z^{-5} \int_0^{z} dz_3 \int_0^{z_3} dz_2 \int_0^{z_2} dz_1 
	f^{(3,2,2)}_3 (z_3 , z_2 , z_1 ) \frac{1}{z_{21}^{4 -2 \epsilon}} \, , \nonumber \\
	H^{(3,3)}_{2} (z)=  - \frac{5 c^2 }{12}  z^{-5} \int_0^{z} dz_3 \int_0^{z_3} dz_2 \int_0^{z_2} dz_1 
	f^{(2,3,2)}_3 (z_3 , z_2 , z_1 ) \frac{1}{z_{31}^{4 -2 \epsilon}} \, , \\
	H^{(3,3)}_{3} (z)=  - \frac{5 c^2 }{12} z^{-5} \int_0^{z} dz_3 \int_0^{z_3} dz_2 \int_0^{z_2} dz_1 
	f^{(2,2,3)}_3 (z_3 , z_2 , z_1 ) \frac{1}{z_{32}^{4 - 2 \epsilon}} \, , \nonumber  
\end{align}
which correspond to diagrams (c), (d), (e) in figure~\ref{Wilson6}.
Integrating over the variables $z_1,z_2,z_3$, we find
\begin{align}
	&H^{(3,3)}_{1} (z)= H^{(3,3)}_{3} (z)=  - \frac{5 c^2 }{12} \cdot  \left[\frac{2 }{45 \epsilon}+\frac{4}{225} (5 \log (z)-1)\right]  \, , \nonumber  \\
	&H^{(3,3)}_{2} (z)=- \frac{5 c^2 }{12} \cdot  \left[ \frac{1}{45 \epsilon}+\frac{2}{675}  (15 \log (z)-26) \right] \, .
\end{align}
Furthermore, we need to consider a contribution of the form as
\begin{align}
	\int_0^z d z_3 \int_0^{z_3} d z_2 \int_0^{z_2} dz_1   f_3 ^{(3,3,3)} ( z_3,z_2 ,z_1)  \langle J^{(3)}  (z_3)  J^{(3)}  (z_2) J^{(3)}  (z_1) J^{(3)} (y) \rangle 
\end{align}
with
\begin{align}
	&\langle  J^{(3)}  (z_3)J^{(3)}  (z_2)J^{(3)}  (z_1) J^{(3)} (y)  \rangle  \nonumber  \\ &  =
	\frac{(5c/6)^2}{z_{32}^{6 - 2 \epsilon} (z_1 - y )^{6 - 2 \epsilon} }   + 
	\frac{(5c/6)^2}{ z_{31}^{6 - 2 \epsilon }  ( z_2 - y )^{ 6 - 2 \epsilon  }} + 	\frac{(5c/6)^2}{z_{21}^{6 - 2 \epsilon} (z_3  - y)^{ 6 - 2 \epsilon} }  + \mathcal{O}(c)\, .
\end{align}
Denoting
\begin{align}
	H^{(3,3)}_{ij} (z) = \left(\frac{5c}{6} \right)^2 z^{-5}
	\int_0^z d z_3 \int_0^{z_3} d z_2 \int_0^{z_2} dz_1   f_3 ^{(3,3,3)} ( z_3,z_2 ,z_1)  
	\frac{1}{z_{ji}^{6 - 2\epsilon}  } \, ,
\end{align}
we find
\begin{align}
	H^{(3,3)}_{12} (z)& =H^{(3,3)}_{23} (z) =  \left(\frac{5c}{6} \right)^2 \left[-\frac{2 }{45 \epsilon} + \frac{1}{225} (9-20 \log (z)) \right] \, ,\nonumber   \\
	H^{(3,3)}_{13} (z)& = \left(\frac{5c}{6} \right)^2 \left[ \frac{1}{225 \epsilon} + \frac{1}{6750}(60 \log (z)-137)  \right] \, . 
\end{align}
Here $H^{(3,3)}_{12} (z)$,  $H^{(3,3)}_{13} (z)$, $H^{(3,3)}_{23} (z)$ correspond to diagrams (f), (g), (h) in figure~\ref{Wilson6}.

Combining the results so far as
\begin{align}
	& z^{ - 5} \lim_{y \to - \infty} |y|^{6 - 2 \epsilon }	\langle W_{-1} (z) J^{(3)}(y) \rangle  \\
	& \qquad = \frac{1}{3} - \left( \frac{6}{c} \right)^2 H^{(2,3)} - \left( \frac{6}{c} \right)^3
	\left[2 H^{(3,3)}_1 (z) +H^{(3,3)}_2 (z)  + 2 H^{(3,3)}_{21} (z)+  H^{(3,3)}_{31} (z)  \right] + \cdots \, , \nonumber
\end{align}
we find
\begin{align}
	z^{-5 } \lim_{y \to - \infty} |y|^{6 - 2 \epsilon }	\langle W_{-1} (z) J^{(3)}(y) \rangle  &= \frac{1}{3} +
	\frac{1}{c}  \left( -\frac{4}{3 \epsilon}+\frac{64 \log (z)}{3}+\frac{1067}{45}  \right) + \cdots 
	\nonumber \\ &= z^{-2} \left[  \frac{1}{3} +
	\frac{1}{ c} \left( - \frac{12}{\epsilon} + \frac{657}{45} \right) \right] \langle W_{-1} (z) \rangle   + \cdots \, .
\end{align}
We remove the divergent term by properly choosing the parameter $c_3$ in \eqref{Wilsonreg} as before. 
We propose to use
\begin{align}
	c_3 = 1 + \frac{1}{c} \left(  \frac{36}{\epsilon} + 1 \right) + \mathcal{O} (c^{-2}) \, ,
	\label{level2c33}
\end{align}
which leads to an extra contribution at the $1/c$ order as
\begin{align}
	\left. z^{-5} \lim_{y \to - \infty} |y|^{6 - 2\epsilon } \frac{6 c_3}{c} \langle W^{(1)'}_{-1} (z) J^{(3)} (y) \rangle \right|_{\mathcal{O} (c^{-1})} = \frac{1}{c} \frac{1}{3} \left(  \frac{36}{\epsilon} + 1 \right) \, .
\end{align}
Here $W^{(1)'}_{-1} (z)$ is defined as
\begin{align}
	W^{(1)'}_{h_0} = \int_0^z d z_1 f^{(3)}_1 (z_1) J^{(3)} (z_1) \, .
	\label{W1p}
\end{align}
Including the effect, we obtain 
\begin{align}
	\lim_{y \to - \infty} |y|^{6 - 2 \epsilon }	\langle W_{-1} (z) J^{(3)}(y) \rangle  =   \frac{1}{3} z^3 \left[ 1 +
	\frac{1}{ c} \frac{224}{5}  \right] \langle W_{-1} (z) \rangle   + \mathcal{O} (c^{-2})
\end{align}
for $\epsilon \to 0$ as in \eqref{3pts3N3}.

\subsection{Two point function at $1/c^2$ order}
\label{2ptN32}

As for $N=2$, we examine the two point function up to the $1/c^2$ order and see whether we can reproduce the $1/c$ correction of conformal weight as in \eqref{dimshift3} after adopting the regularization.
As before, we first evaluate the $1/c$ correction without renormalization and then include its effects.

There are contributions involving only spin two currents, which were already evaluated in \eqref{G2z}. We find
\begin{align}
	c^2	z^{-2} G^{(2)}_\text{spin 2} (z) =  72 \log (z) (4 \log (z)+7)
\end{align}
by setting $h_0 = -1$. In this subsection, we only keep the terms involving $\log (z)$ or $\log ^2 (z)$ and not vanishing at $\epsilon \to 0$.

\begin{figure}
	\centering
	\includegraphics[keepaspectratio, scale=0.7]
	{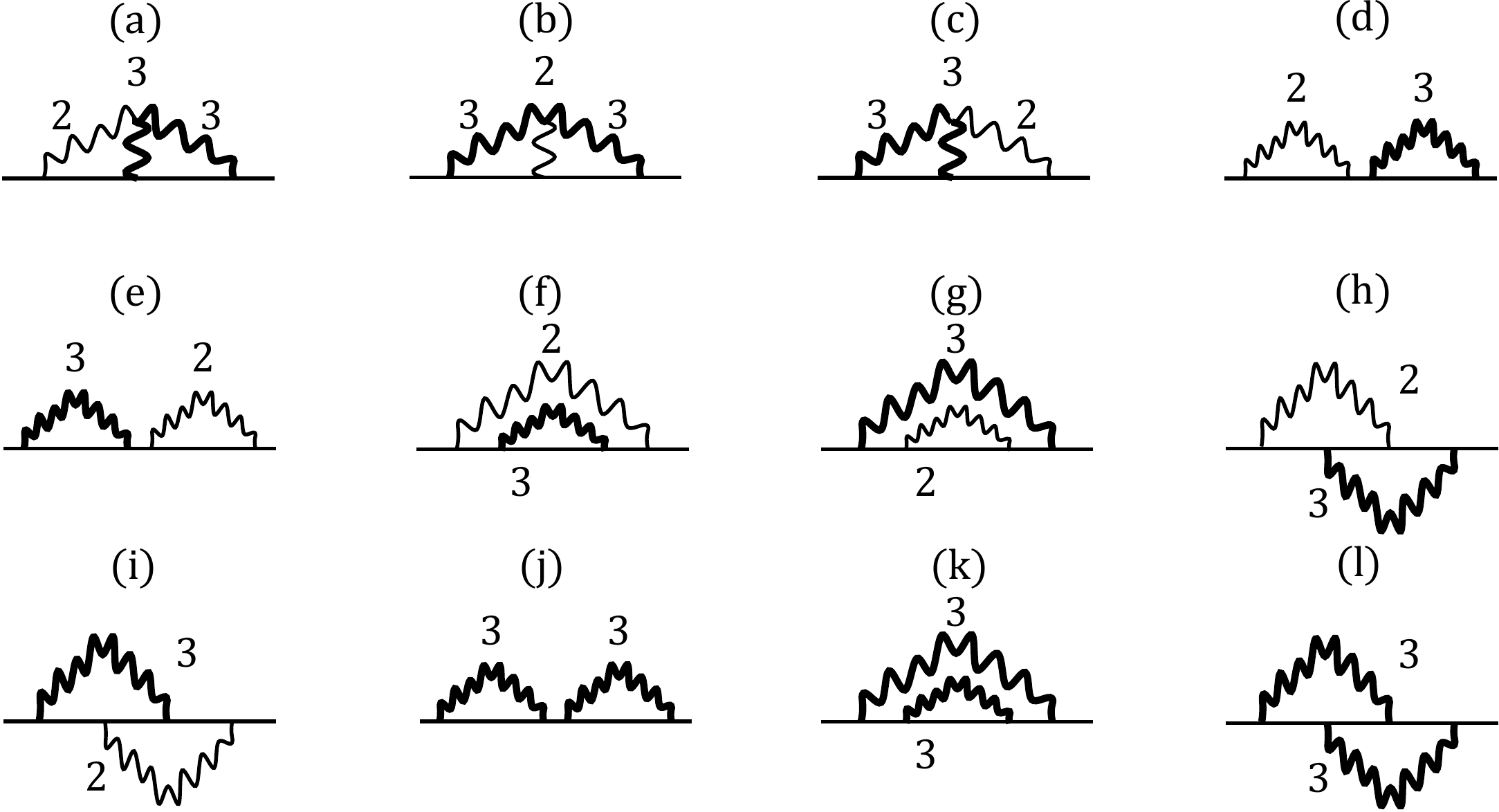}
	\caption{Diagrams contributing to the $1/c^2$ order correction of $\langle \mathcal{O}_{h_+} \bar{\mathcal{O}}_{h_+} \rangle $  for $N=3$ in addition to the ones in figure~\ref{Wilson3}.}
	\label{Wilson7}
\end{figure}

Furthermore, we include the effects of spin three current $J^{(3)}$.
In order to make our notation simpler, we adopt the following rule.
If $J^{(2)} (z_i)$ comes from the open Wilson line, then we use  index $i$. If $J^{(3)} (z_i)$ enters instead of $J^{(2)} (z_i)$, then we replace the index $i$ by {\boldmath{$i$}}. 
We first compute those with three currents  as
\begin{align}
	G^{(2)}_{1\text{\boldmath $23$}} (z)  = & \left( \frac{6}{c}\right)^3 \int_0^z d z_3 \int_0^{z_3} d z_2 \int_0^{z_2} dz_1   f_3^{(3,3,2)} ( z_3,z_2 ,z_1) 	\frac{- 5 c/2}{ z_{32}^{4- \epsilon} z_{31}^{2-\epsilon}  z_{21}^{2- \epsilon} } \, ,
\end{align}
and $G_{\text{\boldmath $1$}2\text{\boldmath $3$}}^{(2)} (z)$, $G_{\text{\boldmath $12$}3}^{(2)} (z)$,   which are represented by diagrams (a), (b), (c) in figure~\ref{Wilson7}, respectively.
Integrations over $z_i$ yield
\begin{align}
	&c^2 z^{-2}	 G_{1\text{\boldmath $23$}}^{(2)} (z) = c^2	 z^{-2} G_{\text{\boldmath $12$}3}^{(2)} (z) = -\frac{ 2880  \log (z) }{\epsilon}  -  4320 \log (z) (\log (z)+1)  \, ,  \nonumber  \\
	&c^2 z^{-2}G_{\text{\boldmath $1$}2\text{\boldmath $3$}}^{(2)} (z) =-\frac{  2880   \log (z)}{\epsilon}- 1440   \log (z) (3 \log (z)+2) \, . 
\end{align}
There are also contributions with two spin two and two spin three currents such as
\begin{align}
	G^{(2)}_{12;\text{\boldmath $34$}} (z)  = & \left( \frac{6}{c}\right)^4 \int_0^z d z_3 \int_0^{z_3} d z_2 \int_0^{z_2} dz_1   f_4^{(3,3,2,2)} ( z_4,z_3 ,z_2,z_1) 	\frac{- 5 c^2 /12}{z_{43}^{6- 2\epsilon}  z_{21}^{4- 2\epsilon} } \, ,
\end{align}
and so on. They are computed as
\begin{align}
	&c^2 z^{-2} G_{12;\text{\boldmath $34$}}^{(2)} (z)  =c^2 z^{-2} G_{\text{\boldmath $12$;34}}^{(2)} (z)  =  \frac{ 960  \log (z)}{\epsilon}  + 160  \log (z) (12 \log (z)+11) \, ,  \nonumber \\
	&c^2 z^{-2} G_{14;\text{\boldmath $23$}}^{(2)} (z)  =  \frac{480 \log (z)}{\epsilon} + 240 \log (z) \left( 4 \log (z)+ 9\right)\, ,\nonumber  \\
	&c^2 z^{-2} G_{\text{\boldmath $14$};23}^{(2)} (z)  = \frac{480 \log (z)}{\epsilon}+ 240 \log (z) \left( 4 \log (z)+ 5 \right) \, ,  \\
	&c^2 z^{-2} G_{13;\text{\boldmath $24$}}^{(2)} (z) = c^2 G_{\text{\boldmath $13$;24}}^{(2)} (z) = \frac{480 \log (z)}{\epsilon} + 350 \log (z) \left(  \log (z)+ 4 \right)  \, , \nonumber
\end{align}
which correspond to diagrams (d)-(i) in figure~\ref{Wilson7}.
Finally, those with four spin three currents are
\begin{align}
	G^{(2)}_{\text{\boldmath $12$;$34$}} (z)  = & \left( \frac{6}{c}\right)^4 \int_0^z d z_3 \int_0^{z_3} d z_2 \int_0^{z_2} dz_1   f_4^{(3,3,3,3)} (z_4,  z_3,z_2 ,z_1) 	\frac{(5 c/6)^2}{z_{43}^{6- 2 \epsilon} z_{21}^{6- 2\epsilon}  } 
\end{align}
and others with different products of the two point function. They are obtained as
\begin{align}
	&c^2 z^{-2} G_{\text{\boldmath $12$;$34$}}^{(2)} (z)  =  \frac{1600  \log (z)}{ \epsilon} + \frac{3200}{3} \left( \log (z) (3 \log (z)+2) \right) \, , \nonumber \\
	&c^2 z^{-2} G_{\text{\boldmath $14$;$23$}}^{(2)} (z)  = \frac{800 \log (z)}{\epsilon}+1600  \log (z) (\log (z)+1)\, , \\
	&c^2 z^{-2} G_{\text{\boldmath $13$;$24$}}^{(2)} (z) = -\frac{160  \log (z) }{ \epsilon} - \frac{8}{3} \log (z) (120 \log (z)+127) \, ,\nonumber
\end{align}
which are represented in diagrams (j), (k), (l) in figure~\ref{Wilson7}, respectively.
Summing up all contributions we have
\begin{align}
	c^2 z^{-2} G^{(2)} (z) = &
	-\frac{2560  \log (z) }{\epsilon}  -\frac{32}{3} \log (z) (48 \log (z)-193)  \, ,
\end{align}
which includes a non-locally divergent term.

Let us then examine the effects of renormalization. There are two types of $1/c$ order corrections as in \eqref{W2h} before the renormalization. Multiplying the $1/c$ terms due to renormalization, some contributions at the $1/c^2$ order arise. With \eqref{level2c32} and \eqref{wfren3}, the contribution with spin two current becomes
\begin{align}
	&\frac{1}{c}\left[ 2 \left( \frac{36}{ \epsilon} + 13 \right) - \frac{32}{\epsilon} - \frac{82}{3} \right] \left(\frac{6}{c} \right)^2 \langle W^{(2)}_{-1} (z) \rangle \nonumber  \\ 
	& \qquad \qquad \qquad =  \frac{1}{c} \left[\frac{960 z^2 \log (z)}{\epsilon}  + 64 z^2  \log (z) (15 \log (z)+17) \right] \, ,
\end{align}
see \eqref{extra} for the previous case with $N=2$.
The contribution with spin three current is
\begin{align}
	&\frac{1}{c} \left[ 2 \left( \frac{36}{ \epsilon} + 1 \right) - \frac{32}{ \epsilon} - \frac{82}{3} \right] \left(\frac{6}{c} \right)^2 \langle W^{(2)'}_{-1} (z) \rangle  \nonumber  \\ 
	& \qquad \qquad \qquad =  \frac{1}{c} \left[ \frac{1600 z^2  \log (z)}{ \epsilon} +\frac{160}{3} z^2 \log (z) \left(30 \log (z)+ 1 \right) \right] \, ,
\end{align}
where we have used \eqref{level2c33} and \eqref{wfren3}.
Thus the $\log(z)$ and $\log ^2 (z)$ dependent terms in the total contribution are
\begin{align}
	&  c^2 z^{-2} \left. \langle \tilde W_{h_0} (z) \rangle \right|_{\mathcal{O}(c^{0})} =  3200 \log (z) + 2048 \log ^2 (z) \, .
\end{align}
The coefficients in front of  $\log (z)$ and  $\log ^2 ( z)$ are precisely those in \eqref{2ptexp} with \eqref{dimshift3}.
We would like to emphasize again that there is cancellation among non-local divergences.

\section{Conclusion and discussions}
\label{conclusion}

We have examined the two and three point functions \eqref{2&3pt} of the 2d W$_N$ minimal model in $1/c$ expansion from the bulk viewpoint. Extending a previous work of  \cite{Besken:2016ooo} at the leading order in $1/c$, we claim that these correlators can be computed with open Wilson lines in sl$(N)$ Chern-Simons gauge theory as in \eqref{2ptWilson} and \eqref{3ptWilson} even at higher orders in $1/c$. There are divergences associated with loop diagrams in the Wilson line computations, and we have to decide how to deal with them.
We offer to regularize the divergences by renormalizing the overall factor of the open Wilson line and parameters $c_s$ introduced in \eqref{Wilsonreg}. 
The finite parts of $c_s$ are fixed such that three point functions from \eqref{3ptWilson} are consistent with the boundary W$_N$ symmetry. We confirm the validity of our prescription by reproducing the $1/c$ corrections of scalar conformal weight from  \eqref{2ptWilson} including $1/c^2$ order terms.

As concrete examples, we have only examined  Chern-Simons gauge theories based on sl$(N)$ with $N=2,3$.
For $N \geq 4$ we see no major difference even though computations would be quite complicated.
For instance, we can reproduce $h_1$ in \eqref{h012} by evaluating integrals in \eqref{2ptWilson} up to the $1/c$ order and comparing the $1/c$ expansion of the two point function in \eqref{2ptexp}.
We consider the following integral as
\begin{align}
	\langle W_{(1 - N)/2}^{(1,s)} (z) \rangle \equiv - \frac{ (2 s -1) N_s}{6} 
	\int_0^z d z_2 \int_0^{z_2} dz_1 f^{(s,s)}_2 (z_2,z_1) \frac{1}{z_{21}^{2 s- 2 \epsilon }}
\end{align}
with conformal weight $(1-N)/2$.
The term including $\log(z)$ at the order $\epsilon^0$
 is evaluated as
\begin{align}
 \langle W_{(1-N)/2}^{(1,s)} (z) \rangle |_{\log , \epsilon^0}
  = (2 s -1) (N^2-1)  \left(- \frac{N_s}{6} \right) ^2 z^{N-1} \log(z) 
\end{align}
for $s=2,3, \ldots ,10$. We conjecture that the above equality also holds for $s > 10$.
 Then, the $1/c$ order correction of scalar conformal weight for generic $N$  can be read off as
\begin{align}
	- \frac{1}{2}  \sum_{s = 2}^N   \left(- \frac{6}{N_s} \right) ^2  (2 s -1) (N^2-1)  \left(- \frac{N_s}{6} \right) ^2 = - \frac{(N^2 - 1)^2}{2} \, ,
\end{align}
which matches  $h_1$ in \eqref{h012}.
For our purpose it is enough to work with the non-unitary duality, but other problems may require a unitary one, i.e., the 't~Hooft limit of \cite{Gaberdiel:2010pz}, see  footnote \ref{nonunitary}. For the unitary duality,
we should extend the analysis to the case with a higher spin algebra hs$[\lambda]$, which is a gauge algebra of 3d Prokushkin-Vasiliev theory \cite{Prokushkin:1998bq}. In particular, we would like to understand the precise relation between open Wilson lines and particles traveling in the bulk.

An important open problem is to confirm our proposal that correlators in the 2d W$_N$ minimal model can be computed with open Wilson lines in sl$(N)$ Chern-Simons gauge theory including $1/c$ corrections.
In particular, we have to extend the checks to higher orders in $1/c$.
We have conjectured that all divergences are removed by renormalizing  the overall factor of the open Wilson line and the parameters $c_s$ in \eqref{Wilsonreg}, but it is desirable to prove this claim.
A different regulator was introduced in \cite{Besken:2017fsj} by shifting $1/(z_{21}^2)^a \to 1/(z_{21}^2 + \epsilon^2)^a$, but it breaks conformal symmetry.
We can see that divergences from loop computations with this regulator cannot be absorbed by these changes,
thus conformal symmetry in the regularization procedure should play an important role.

We have proposed our regularization prescription so as to be analogous to that for usual quantum field theory even though the precise relation is yet to be clarified.
We offer to fix the interaction parameters by comparing them to ``experimental data'' that are obtained from dual conformal field theory in the current situation. Once they are fixed, then other quantities like the self-energy of the scalar propagator are claimed to be predictable. A particularly nice thing happens for $N=2$. In this case, the $1/c$ order of the interaction parameter $c^{(1)}_2$ was determined by using the information on $h_1$ in \eqref{dimcorr} through \eqref{Ward}. Fortunately, $h_1$ can be obtained from the expectation value of the open Wilson line as in \eqref{2pt1}, therefore we do not need to refer to explicit boundary data and everything is computable in terms of bulk theory. Here we have only considered to the next leading order in $1/c$, but it is natural to expect that the same is true for higher orders in $1/c$ as well.
For $N=3$, we fixed the $1/c$ order of the other interaction parameter $c_3^{(1)}$ such that the equality in \eqref{3pts3N3} is satisfied. Here the number $224/5$ was borrowed from the W$_N$ minimal model. However, we believe that there should be a way to determine $c_3$ without referring to explicit boundary data, and it is an important open problem to find this out.
We do not claim that our prescription is unique, and in fact a different one was adopted in \cite{Fitzpatrick:2016mtp} for $N=2$.  It is easier to see the physical meaning in our regularization procedure, but their prescription seems to be convenient for actual computations of conformal blocks.
In any case, it should be useful to understand the relation between different prescriptions.

In this paper, we have examined the  duality of \cite{Gaberdiel:2010pz} in the semiclassical limit discussed in \cite{Castro:2011iw,Gaberdiel:2012ku,Perlmutter:2012ds} with $1/c$ corrections, but it is also possible to extend the analysis to other examples. In particular, an $\mathcal{N}=2$ supersymmetric version of duality was proposed in \cite{Creutzig:2011fe}, and the bulk description of its semiclassical limit was argued to be given by sl$(N+1|N)$ Chern-Simons gauge theory \cite{Hikida:2012eu}. See  \cite{Tan:2012xi,Datta:2012km,Chen:2013oxa,Datta:2013qja,Banados:2015tft} for conical defect or black hole solutions in higher spin supergravity. 
We think that supersymmetric extension is important for the following two reasons.
Firstly, it is usually expected that supersymmetry suppresses quantum effects, and it would enable us to examine higher order corrections in $1/c$ systematically.
Secondly, supersymmetry helps us to study relations between higher spin gauge theory and superstring theory, and concrete examples have been discussed in \cite{Creutzig:2013tja,Creutzig:2014ula,Hikida:2015nfa} with $\mathcal{N}=3$ supersymmetry and in \cite{Gaberdiel:2013vva,Gaberdiel:2014cha} with $\mathcal{N}=4$ supersymmetry.
We would like to report on this extension in the near future.

\subsection*{Acknowledgements}

We are grateful to Andrea Campoleoni, Pawel Caputa, Nilay Kundu, Takahiro Nishinaka, Volker Schomerus, Yuji Sugawara, Tadashi Takayanagi, and J\"{o}rg Teschner for useful discussions.
YH would like to thank the organizers of the ``Universit\"{a}t
Hamburg-Kyoto University Symposium" and the workshop ``New ideas on higher spin gravity and holography'' at Kyung Hee University, Seoul for their hospitality.
The work of YH is supported by JSPS KAKENHI Grant Number 16H02182.

%

\providecommand{\href}[2]{#2}\begingroup\raggedright\endgroup

\end{document}